\documentclass{aa}  

\usepackage{graphicx,natbib}
\usepackage{txfonts,color}
\usepackage[draft]{hyperref}
\begin{document}

   \title{The environments of radio-loud AGN from the LOFAR Two-Metre Sky Survey (LoTSS)}

   \author{J.H.~Croston
          \inst{1}
          \and
          M.J.~Hardcastle\inst{2}
          \and
          B.~Mingo\inst{1}
          \and
          P.N.~Best\inst{3}
          \and
          J.~Sabater\inst{3}
          \and
          T.M.~Shimwell\inst{4}
          \and
          W.L.~Williams\inst{2}
          \and
          K.J.~Duncan\inst{5}
          \and
          H.J.A.~R\"{o}ttgering\inst{5}
          \and
          M.~Brienza\inst{6}
          \and
          G.~G\"{u}rkan\inst{7}
          \and
          J.~Ineson\inst{8}
          \and
          G.K.~Miley\inst{5}
          \and
          L.M.~Morabito\inst{9}
          \and
          S.P.~O'Sullivan\inst{10}
          \and
          I.~Prandoni\inst{6}
          }

   \institute{School of Physical Sciences, The Open University, Walton Hall, Milton Keynes, MK7 6AA, UK\\
              \email{Judith.Croston@open.ac.uk}
         \and
             Centre for Astrophysics Research, University of Hertfordshire, College Lane, Hatfield AL10 9AB
         \and
         SUPA, Institute for Astronomy, Royal Observatory, Blackford Hill, Edinburgh, EH9 3HJ, UK
         \and
         ASTRON, the Netherlands Institute for Radio Astronomy, Postbus 2,7990 AA, Dwingeloo, The Netherlands
         \and
       Leiden Observatory, Leiden University, PO Box 9513, NL-2300 RA Leiden, the Netherlands
         \and
       INAF - Istituto di Radioastronomia, Via P. Gobetti 101, 40129 Bologna, Italy
       \and
         CSIRO Astronomy and Space Science, PO Box 1130, Bentley WA 6102, Australia
         \and
        School of Physics and Astronomy, University of Southampton, Highfield, Southampton SO17 1BJ, UK
       \and 
       Astrophysics, University of Oxford, Denys Wilkinson Building, Keble Road, Oxford OX1 3RH, UK
       \and
       Hamburger Sternwarte, Universit\"at Hamburg, Gojenbergsweg 112, D-21029 Hamburg, Germany
       }

   \date{}

% \abstract{}{}{}{}{} 
% 5 {} token are mandatory
 
  \abstract
  % context heading (optional)
  % {} leave it empty if necessary  
{An understanding of the relationship between radio-loud active galaxies and their large-scale environments is essential for realistic modelling of radio-galaxy evolution and environmental impact, for understanding AGN triggering and life cycles, and for calibrating galaxy feedback in cosmological models. We use the LOFAR Two-Metre Sky Survey (LoTSS) Data Release 1 catalogues to investigate this relationship. We cross-matched a sample of 8,745 radio-loud AGN with $0.08<z<0.4$, selected from LoTSS, with two Sloan Digital Sky Survey (SDSS) cluster catalogues, and find that only 10 percent of LoTSS AGN in this redshift range have a high-probability association, so that the majority of low-redshift AGN (including a substantial fraction of the most radio-luminous objects) must inhabit haloes with $M < 10^{14}$ M$_{\sun}$. We find that the probability of a cluster association, and the richness of the associated cluster, is correlated with AGN radio luminosity, and we also find that, for the cluster population, the number of associated AGN and the radio luminosity of the brightest associated AGN is richness-dependent. We demonstrate that these relations are not driven solely by host-galaxy stellar mass, supporting models in which large-scale environment is influential in driving AGN jet activity in the local Universe. At the lowest radio luminosities we find that the minority of objects with a cluster association are located at larger mean cluster-centre distances than more luminous AGN, an effect that appears to be driven primarily by host-galaxy mass. Finally, we also find that FRI radio galaxies inhabit systematically richer environments than FRIIs, consistent with previous work. The work presented here demonstrates the potential of LoTSS for AGN environmental studies. In future, the full northern-sky LoTSS catalogue, together with the use of deeper optical/IR imaging data and spectroscopic follow-up with WEAVE-LOFAR, will provide opportunities to extend this type of work to much larger samples and higher redshifts.}

   \maketitle
%
%________________________________________________________________

\section{Introduction}

The large-scale environments of radio galaxies have been the subject of research for more than forty years, and a link between radio galaxies and galaxy clusters is well established \citep[e.g.][]{longair79,prestage88}. For the well-studied powerful radio galaxies it has long been argued that group or cluster-like external pressures are required in order to provide a medium to confine the expanding radio-lobe plasma, while in the local Universe it is thought that accretion from material originating in the hot group/cluster atmospheres of massive galaxies \citep[e.g.][]{gaspari13,gaspari17} fuels their radio-loud AGN activity and the associated feedback cycle required to regulate their star formation, and that massive black holes such as those found in brightest cluster galaxies may be a requirement for powerful jet activity. Luminous high-redshift radio galaxies at $z>2$ are often associated with rich protoclusters, indicating that in the early Universe powerful jets typically inhabit the highest density regions, where the most massive galaxies are forming \citep[e.g.][]{miley08}. 

A range of methods have been applied to examine the environments of radio galaxies \citep[e.g.][]{morganti88,prestage88,hill91,zirbel97,worrall00,best04,croston08,tasse08,gendre13,sabater13,ineson15,osullivan15,magliocchetti18}. The best-determined environmental measurements come from X-ray observations, which provide a good proxy for group/cluster mass while enabling direct measurement of the external pressure profile into which the radio lobes expand; however, the deep X-ray observations required to measure group properties cannot realistically be obtained for poor environments at moderate redshifts, or for very large samples, with current X-ray facilities. Optical and infrared environmental measures provide a less direct proxy for cluster mass and do not provide direct information about external pressure distributions, but they do provide useful statistical information for the population, and are now readily available for the local Universe from survey data over large sky areas. 
  
There are several motivations for determining the relationship between radio galaxies and group/cluster environments; these include investigating the triggering of AGN activity \citep[e.g.][]{sabater13,sabater18}, determining when and where particular modes of AGN feedback occur \citep[e.g.][]{ineson13,ineson15}, determining physical conditions in radio-galaxy jets and lobes \citep[e.g.][]{croston08,ineson17,croston18}, testing models of source dynamics and environmental impact \citep[e.g.][]{hardcastle18a}, and developing the use of radio galaxies as a way to locate clusters and groups at the highest redshifts \citep[e.g.][]{osullivan15,croston17,johnstonh15}. Much work has focused on examining the host galaxies of AGN in the interests of understanding the mechanisms driving triggering and feedback, but for a substantial fraction of the population it is on the group/cluster scale that much of the jet energy is injected, and so the links between radio properties and group/cluster environment are also of strong interest, independent of host galaxy properties. 

The two main approaches to exploring these links have been (1) to consider a radio-selected sample of AGN and then investigate their environments, or (2) to consider a sample of galaxy clusters and investigate their radio-loud AGN properties. The overwhelming conclusion of studies of the first type is that, while some radio galaxies are found in rich clusters, the preferred environments of radio-loud AGN in the nearby Universe are moderately rich environments, i.e. galaxy groups \citep[e.g.][]{best04,worrall00,croston08,ineson15,ching17}. It is also now well established from studies of the second type that galaxy clusters have a very high probability of hosting a central radio-loud AGN -- approaching 100 per cent probability for the most massive brightest cluster galaxies \citep{best07}. 

Recent work based on modern optical surveys \citep[e.g.][]{best04,sabater13} has improved on the limitations of small samples and narrow radio luminosity ranges of earlier environmental studies, while recent X-ray studies have better constrained the link between radio properties and cluster richness \citep{ineson15}. However, there remain outstanding questions that can be addressed by combining large radio samples with the most recent group/cluster catalogues. A crucial question in the context of understanding AGN feedback is how the jet properties that control the quantity and locations of energy injection (e.g. jet powers, source lifetimes and duty cycles) relate to group/cluster properties.  Specifically, for a jet of a given power that extends to sizes greater than a few tens of kpc, the locations of energy deposition, the mechanisms (e.g. the relative contributions of strong and weak shocks, sound wave distribution of energy across the ICM volume), and the relative energetic importance of the jet's energy input to the thermal energy of the intragroup or intracluster medium, will all depend on the richness of the large-scale environment. A number of studies have identified a relationship between radio luminosity and environmental richness \citep[e.g.][]{best04,ineson13,ineson15,ching17}, with richer environments appearing to host more luminous radio galaxies. Most recently we have shown that this relationship may be linked both to radio-galaxy accretion mode \citep{ineson15} and to source morphology \citep{croston18}. It is important to place these relationships between AGN properties and environment on a firmer footing as we move to the era of large, deep, wide-field radio surveys, to enable robust inference of the population-wide impact of radio-loud AGN on galaxy evolution.

Additional motivation for understanding the relationship between AGN properties and group/cluster environment comes from the study of diffuse radio sources in clusters (radio relics, halos and mini-halos). It has been recognised for some time that a seed population of relativistic particles is required by the favoured models to explain these extended radio structures \citep[e.g.][]{brunetti14}, and radio-loud AGN provide an obvious source of these particles. Late-stage radio-lobe evolution and the mixing of radio-lobe plasma into the ICM is also likely to be important for cluster magnetic-field evolution \citep[e.g.][]{xu2010}. Low-frequency radio observations are beginning to confirm the important connections between radio galaxies and other extended cluster radio sources \citep[e.g.][]{bonafede14,vanweeren17}. Understanding the connections between jet activity and environment, including, for example, the prevalence and properties of radio-loud AGN in cluster outskirts, is important in this context.

In this paper, we carry out the first environmental investigation for the newly released LoTSS DR1 catalogue for the HETDEX Spring Field \citep{shimwell18}. LOFAR-DR1 improves on previous radio surveys in several ways: it is more than an order of magnitude deeper than FIRST \citep{becker95} for sources of typical spectral index, its effective survey frequency of 144 MHz\footnote{For consistency with other LoTSS publications, and simplicity of comparison with other surveys, we refer to the survey frequency as 150 MHz in the remainder of the paper. The small frequency offset does not have any significant effect on plotted quantities.} makes it particularly sensitive to steep spectrum emission, such as remnant and restarting sources \citep[e.g.][]{brienza17,mahatma18}, and LOFAR's {\it uv} coverage provides both good spatial resolution (6 arcsec) and sensitivity to extended structure, improving estimates of source properties, including size and luminosity, and enabling reliable morphological characterisation for extended sources. At $z<0.4$, the redshift range of interest to this study, the survey is flux-complete down to the relatively low 150-MHz luminosity of $\sim 3 \times 10^{23}$ W Hz$^{-1}$.  Here we make use of two Sloan Digital Sky Survey (SDSS) group/cluster catalogues with well-calibrated richness estimators -- the DR8 RedMaPPer catalogue \citep{rykoff14} and the DR8 photo-$z$ cluster catalogue of \citet{wen2012} -- to obtain the largest sample of cross-matched radio-galaxy/cluster associations studied to date. In Section~\ref{sec:sample} we provide more detail about the parent catalogues and sample selection, and about our cross-matching procedure, in Section~\ref{sec:results} we present an analysis (i) of the environmental properties of our low-frequency selected AGN sample, and (ii) of the low-frequency AGN properties of our group/cluster samples, and in Section~\ref{sec:discussion} we discuss the implications of our results.
\begin{figure}
   \includegraphics[width=8.8cm]{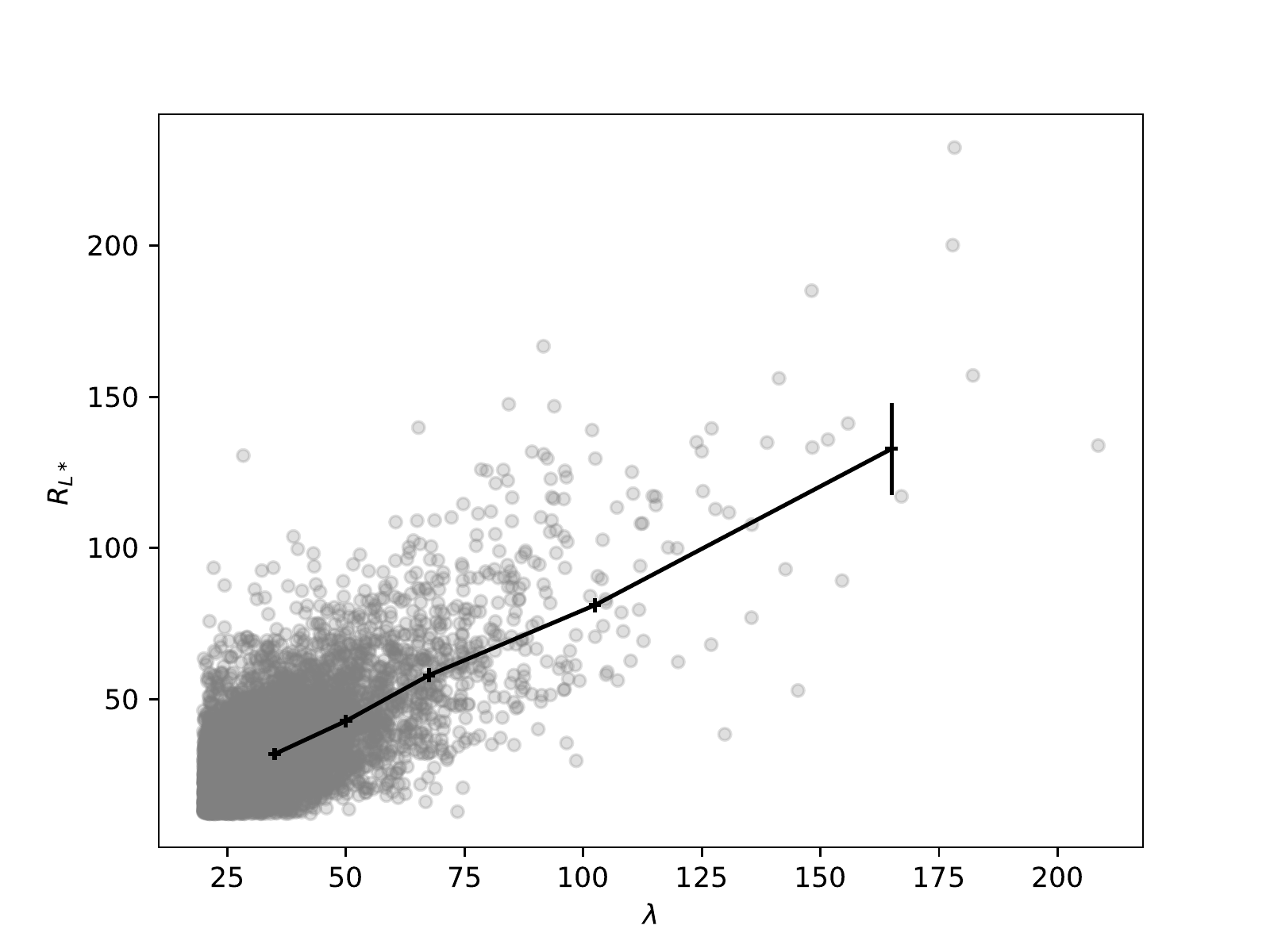}
\caption{Comparison of the two SDSS richness estimators used in this paper, RedMaPPer $\lambda$ and \citet{wen2012} R$_{L*}$, for all clusters from the two catalogues matched using the same method as for our analysis (grey points). The solid black line shows the mean value of $R_{L*}$ in bins of $\lambda$.}
\label{fig:richcompare}
\end{figure}

\begin{figure*}
  \centering
  \includegraphics[width=8.8cm]{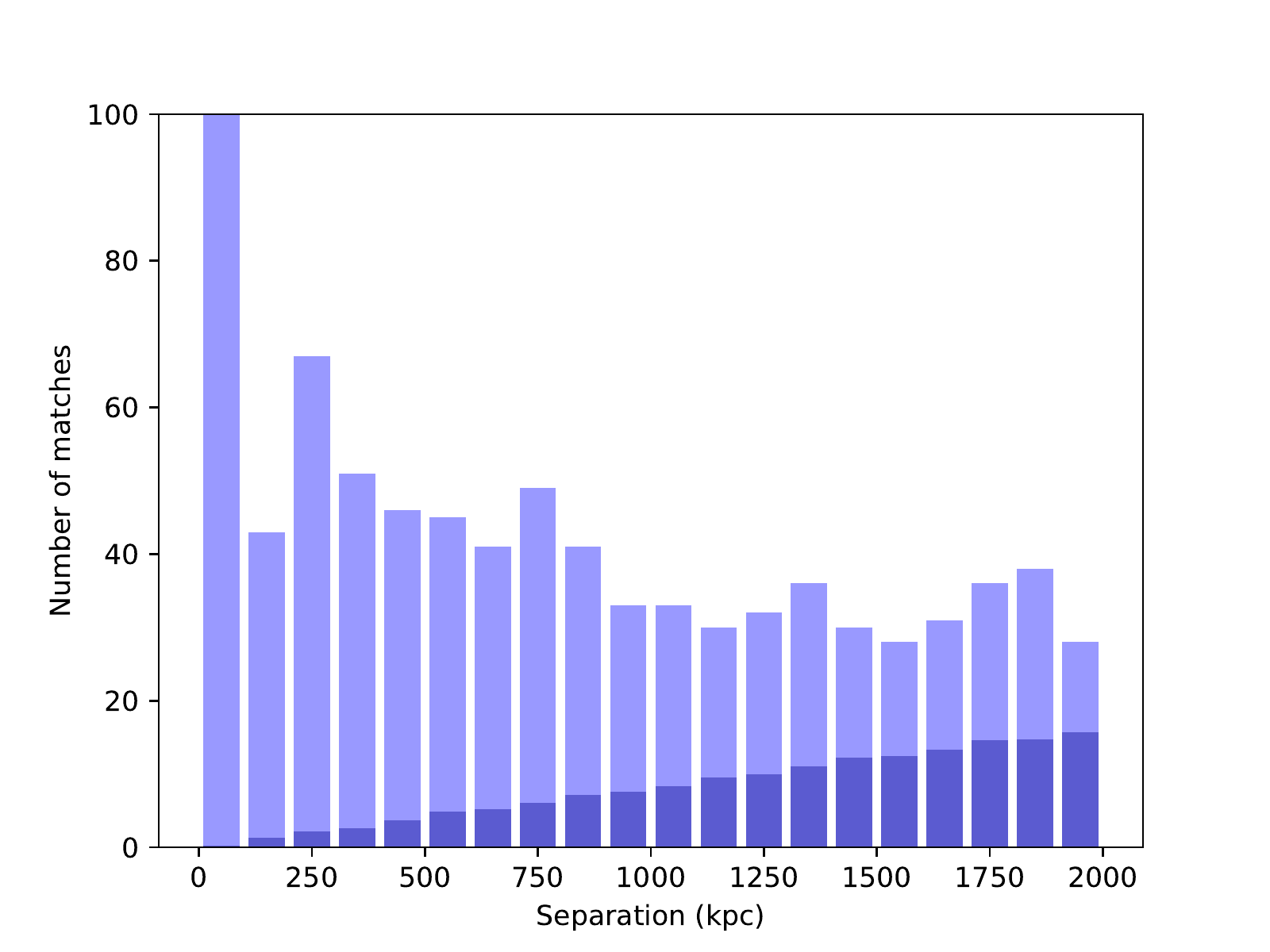}
\includegraphics[width=8.8cm]{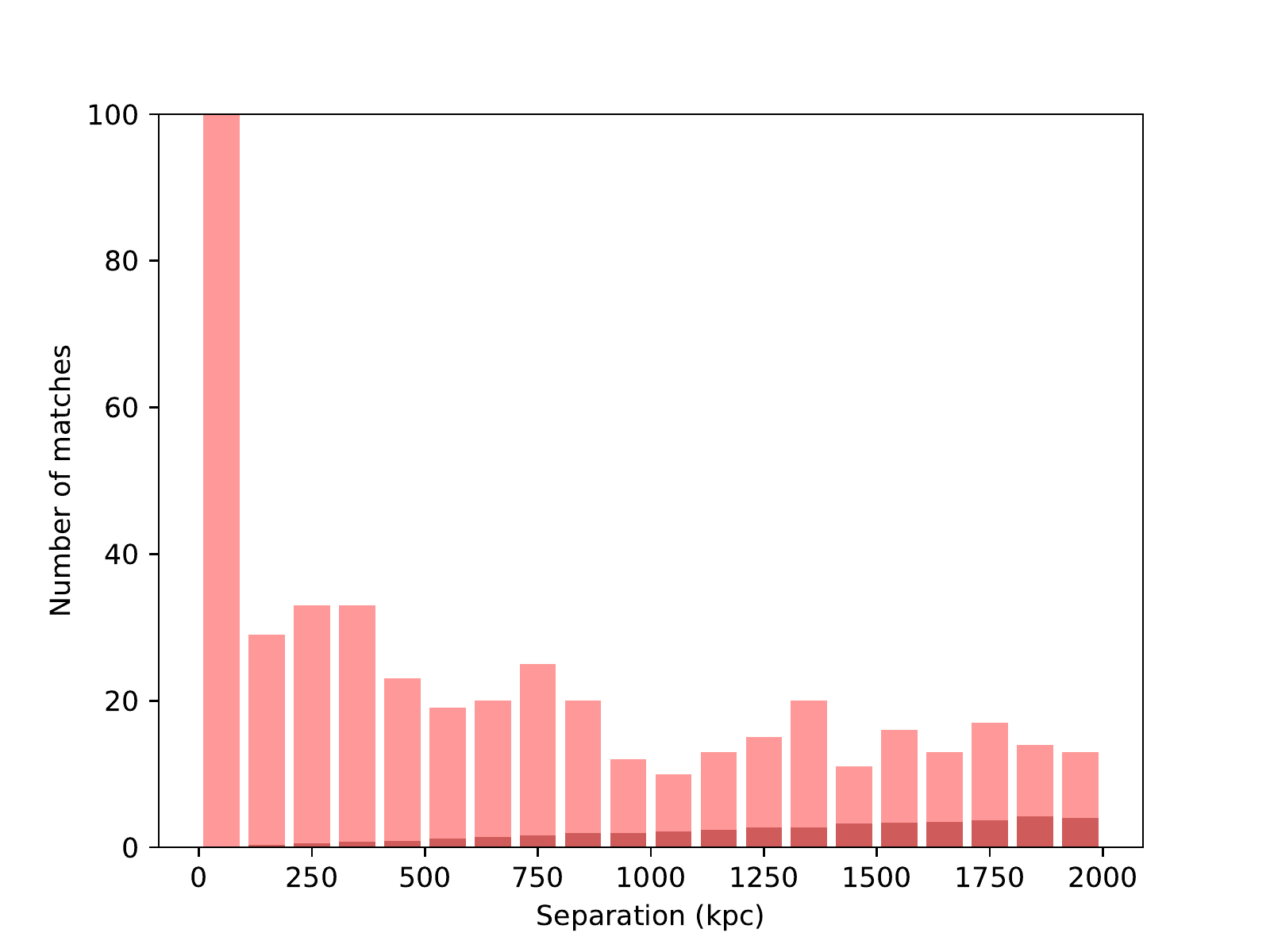}
  \caption{Separation distribution for cross-matched catalogue and for spurious matches from cross-matching with randomized catalogues for the \citet{wen2012} catalogue (left; blue and dark blue) and \citet{rykoff14} catalogue (right; red and dark red). Note that the y axis is cropped to show the distribution at large distances: the bin encompassing zero separation in both catalogues extends to much higher values (630 for the left-hand and 232 for the right-hand panel) -- these are sources for which the AGN host and cluster centre galaxy are the same. }
              \label{fig:cats}%
\end{figure*}
  
\section{Sample and methods}
\label{sec:sample}
The LoTSS-DR1 catalogue \citep{shimwell18} contains 325,694 sources over 424 deg$^{2}$, of which 73 percent have optical identifications \citep{williams18}, and 70 percent of the identified sources (51 percent of the full catalogue) have either spectroscopic or photometric redshifts \citep{duncan18}. Separating AGN from sources in which the radio emission is dominated by star formation processes is a challenge for LoTSS (as for all deep radio surveys), and this is discussed by \citet{sabater18} and \citet{hardcastle18b}. For the work reported here we use the full radio-loud AGN sample described by \citet{hardcastle18b}, which extends the spectroscopic AGN sample of \citet{sabater18}, and consists of the subset of the value-added LoTSS-DR1 catalogue of \citet{williams18} that have optical/IR IDs, robust redshifts, and meet the radio-loud AGN (RLAGN) criteria described by \citet{hardcastle18b} (incorporating spectroscopic criteria where present, as well as using additional criteria based on mid-infrared, optical and radio properties). This RLAGN sample contains 23,344 objects.

\subsection{Environmental datasets}

Our aim here is to carry out a preliminary environmental analysis of the LOFAR radio-loud AGN population. In the long term we intend to carry out in-depth environmental analysis using PanSTARRS, WISE, and deeper spectroscopic data, so as to fully exploit the depth of the LOFAR DR1 dataset and extend this work to higher redshifts; however, here we make use of existing cluster catalogues from the Sloan Digital Sky Survey (SDSS), the RedMaPPer catalogue of \citet{rykoff14}, and the \citet{wen2012} group/cluster catalogue, both of which are based on SDSS DR8. We choose to use two different catalogues constructed with different methods with the aim of drawing conclusions that are relatively robust to the cluster finding and richness estimation methods (although selection biases will nevertheless be present as both methods are based around the properties of the brightest cluster member galaxies). The \citet{rykoff14} RedMaPPer catalogue (hereafter the R14 catalogue) contains $\sim 1,000$ clusters over the LOFAR HETDEX survey footprint, selected using a red-sequence finding method optimised to minimise scatter on the mass-richness relation. The \citet{wen2012} group/cluster catalogue (hereafter the W12 catalogue) extends to somewhat lower richness, with $\sim 4000$ clusters over our survey area, and is based on an iterative method incorporating photometric redshift selection and a friends-of-friends method. Both methods have comparatively well-calibrated richness estimators. The R14 $\lambda$ estimator is a sum of the probabilities of membership for candidate cluster members obtained from red-sequence modelling, and \citet{rozo14} report a scatter in mass at fixed $\lambda$ of $\sim 25$ percent. For the W12 catalogue we use their R$_{L*}$ richness estimator, which is based on the overdensity of optical luminosity (L$_{200}$/L$_{*}$), and for which they estimate a scatter in mass at fixed R$_{L*}$ of $\sim 21$ per cent. In Fig.~\ref{fig:richcompare} we compare the two cluster richness measures for objects with matches in both catalogues. The two measures are well correlated, which is reassuring; however, there is a fairly large scatter which is not unexpected for optical richness measures in the relevant redshift range. The relation between mean $R_{L*}$ and mean $\lambda$ is offset from a 1:1 line, with $\lambda$ values typically systematically higher than $R_{L*}$ for the same cluster -- such a difference is not surprising given the different richness definitions, but should be borne in mind in the analysis that follows. It is evident that environmental richness estimates for individual AGN will have large uncertainties, but our aim in this work is to investigate the statistical properties of the population as a whole. 

Based on the completeness of the two environmental catalogues, we impose a redshift range of $0.08 < z <0.4$ for our environmental comparisons. In this redshift range the W12 catalogue is $>95$ percent complete above M$_{200}> 10^{14}$ M$_{\sun}$ to $z=0.42$), while the R14 catalogue is $>85$ percent complete above $\lambda=30$ and $>95$ percent above $\lambda=40$. Our imposed redshift selection reduces the LoTSS-DR1 RLAGN sample to 8,745 objects. In this redshift range the AGN sample is flux-complete above $L_{150} \sim 3 \times 10^{23}$ W Hz$^{-1}$. 52 per cent of the AGN have spectroscopic redshifts, with the remainder having well-constrained photometric redshifts. In constructing the parent AGN sample, \citet{hardcastle18b} discarded a subset of optically identified sources for which the photometric IDs were poorly constrained before carrying out separation of AGN and star-forming galaxies. Assuming that the fraction of discarded (poorly constrained) photometric redshifts is similar for AGN and star-forming galaxies, then in our redshift range of $0.08<z<0.4$ we may be missing up to $12\%$ of the RLAGN population -- this is a conservative upper limit, as it is expected that the star-forming galaxies will typically have fainter hosts at a given redshift, and therefore are more likely to be discarded. The discarded objects in this redshift range have a similar distribution of radio luminosities to those retained, but their host galaxy rest-frame $K_{S}$ magnitudes are on average half a magnitude fainter. Any effect from this incompleteness is likely to be small, and does not affect our later conclusions. Over the sample redshift range the photometric redshifts determined by \citet{duncan18} have a typical uncertainty of $\sigma = 0.03$ and an outlier fraction of $1.1 \%$ for the objects meeting the ``good'' photo-$z$ criterion described above. We choose to use the full radio luminosity range available, and not to restrict ourselves to the spectroscopic sample, so as to get as complete a view of the population as possible and to maximise the information we can obtain, but we carry out careful checks to ensure redshift- and host galaxy-dependent selection effects in either the AGN or cluster catalogues do not influence our conclusions.

\subsection{Cross-matching of the AGN and cluster catalogues}

We cross-matched the LOFAR-DR1 sample of 8,745 AGN separately with each cluster catalogue, using a combination of projected physical distance at the redshift of the AGN ($\Delta D$) and redshift offset ($\Delta z$). We have chosen to use fixed matching thresholds in $\Delta D$ and $\Delta z$, rather than cluster-mass-dependent thresholds, because of the large uncertainities on the richness estimates for individual clusters, but we have also tested the effect of using mass-dependent matching thresholds, as explained below.

In order to determine the optimal search radius to identify cluster candidates, we initially used a maximum radius of 2 Mpc at the redshift of each AGN and a maximum $\Delta z$ of 0.01 (corresponding to $\sim 3$ times the typical velocity dispersion of a massive cluster). We then used randomization of the catalogued cluster positions to investigate how the fraction of spurious associations depends on search radius. Fig.~\ref{fig:cats} shows the distribution of $\Delta D$ for matches with the two cluster catalogues compared with the mean distribution for spurious matches obtained from 100 trials of cross-matching with a randomised version of the cluster catalogue. We find that a maximum search radius of 1 Mpc offers a good compromise, in which the total fraction of spurious matches is $<2$ per cent for R14 and $<3$ per cent for W12. The highest spurious fraction is found above 800 kpc for the W12 matches, but remains $< 25$ per cent in the range 800 kpc to 1 Mpc. We also tested the effects of varying the redshift threshold and the use of a fractional rather than an absolute threshold in redshift, which did not significantly affect the results. We note that the paper's conclusions (Section~\ref{sec:conclusions}) are also robust to a choice of cross-matching search radius between 500 kpc and 2 Mpc. 

Both our AGN sample and the two cluster catalogues include objects with spectroscopic redshifts as well as those with only photometric redshifts, whose redshift uncertainty needs to be taken into account in the matching process. For each AGN we therefore determined an association probability for any group/cluster within the matching radius of 1 Mpc, defined as the probability that the redshift separation between the AGN and cluster is less than $\Delta z=0.01$, assuming a Gaussian probability distribution for the redshifts of the AGN and cluster based on their reported uncertainties. We then compiled an AGN environment catalogue consisting of the most probable cluster match for each AGN, if one exists, and separately compiled two cluster catalogues incorporating the number of AGN matches for each cluster, the 150-MHz luminosity of the brightest associated AGN, the physical size of the brightest associated AGN, and the physical size of the largest associated AGN. 

Finally, we also investigated the effect of using mass-dependent matching thresholds instead of fixed values for the entire sample, scaling $\Delta D$ and $\Delta z$ according to the dependence of $R_{200}$ and velocity dispersion on richness, respectively. We investigated the effect of using a mass-scaled radius together with a fixed redshift threshold, and of using mass-scaled values for both thresholds - this reduced the match fraction by 30 percent, but the distributions of $L_{150}$ and of the two richness measures, $R_{L*}$ and $\lambda$, is unchanged. Hence, if our choice of fixed matching thresholds is leading to spurious matches, this does not appear to be having a systematic effect on the matched sample properties. We therefore choose to adopt the original fixed matching thresholds of $\Delta D = 1$ Mpc and $\Delta z = 0.01$, so as to avoid introducing noise into the cross-matching statistics from the poorly constrained individual richness measurements.

In the analysis that follows we make use of the matching probabilities as follows: when considering the properties of the AGN sample, we aim to make use of the full statistical information contained in the matching probabilities, and so incorporate all cluster matches to determine weighted mean properties for the AGN environments. However, for our analysis of the properties of the full cluster samples, in order to reliably identify the brightest and largest cluster AGN, we considered only AGN with an association probability above 0.8. Our three cross-matched catalogues form the basis of the analysis presented in the following section. 

\section{Results}
\label{sec:results}

\begin{figure*}
   \centering
   \includegraphics[width=9cm]{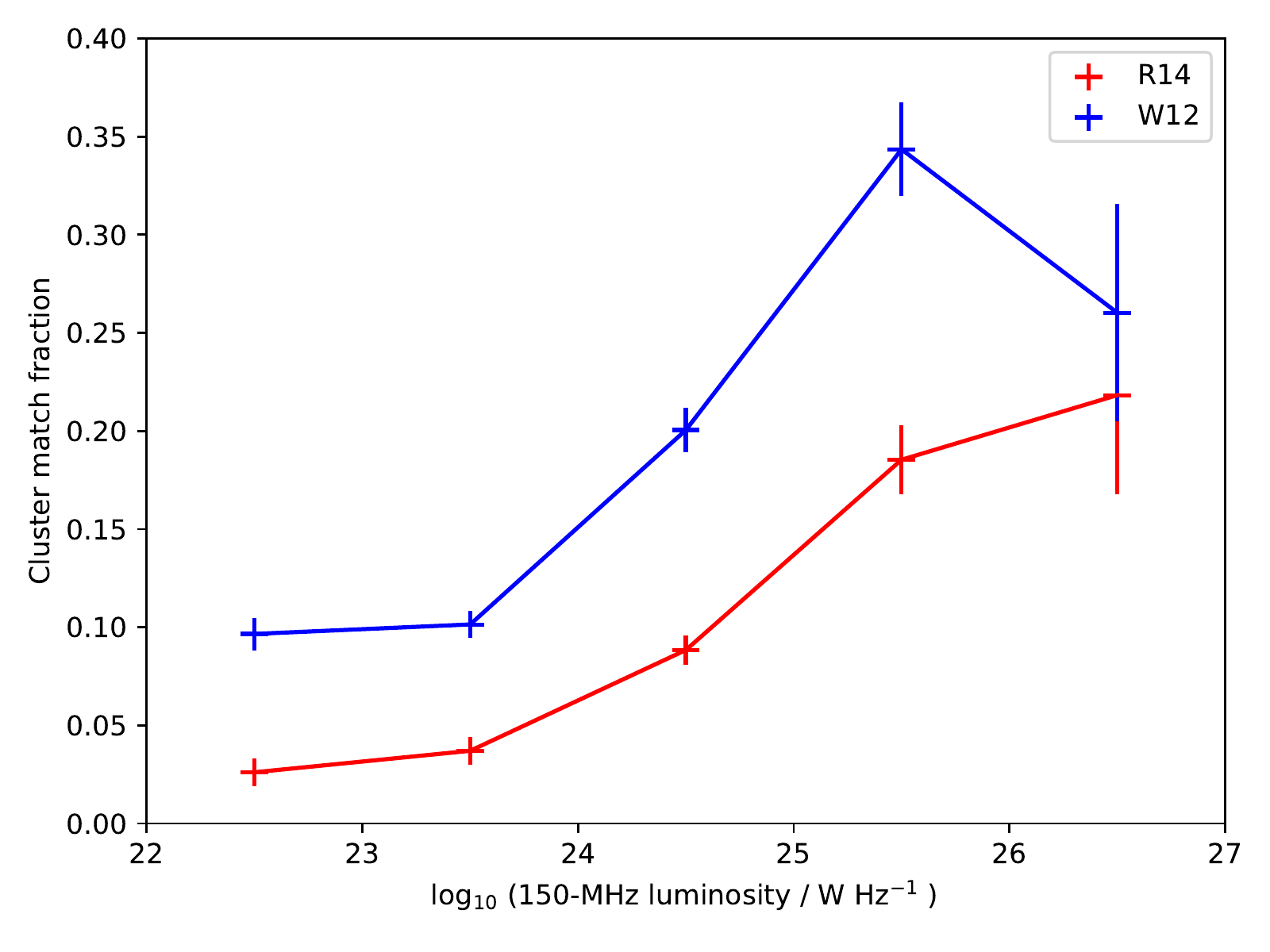}	
   \includegraphics[width=9cm]{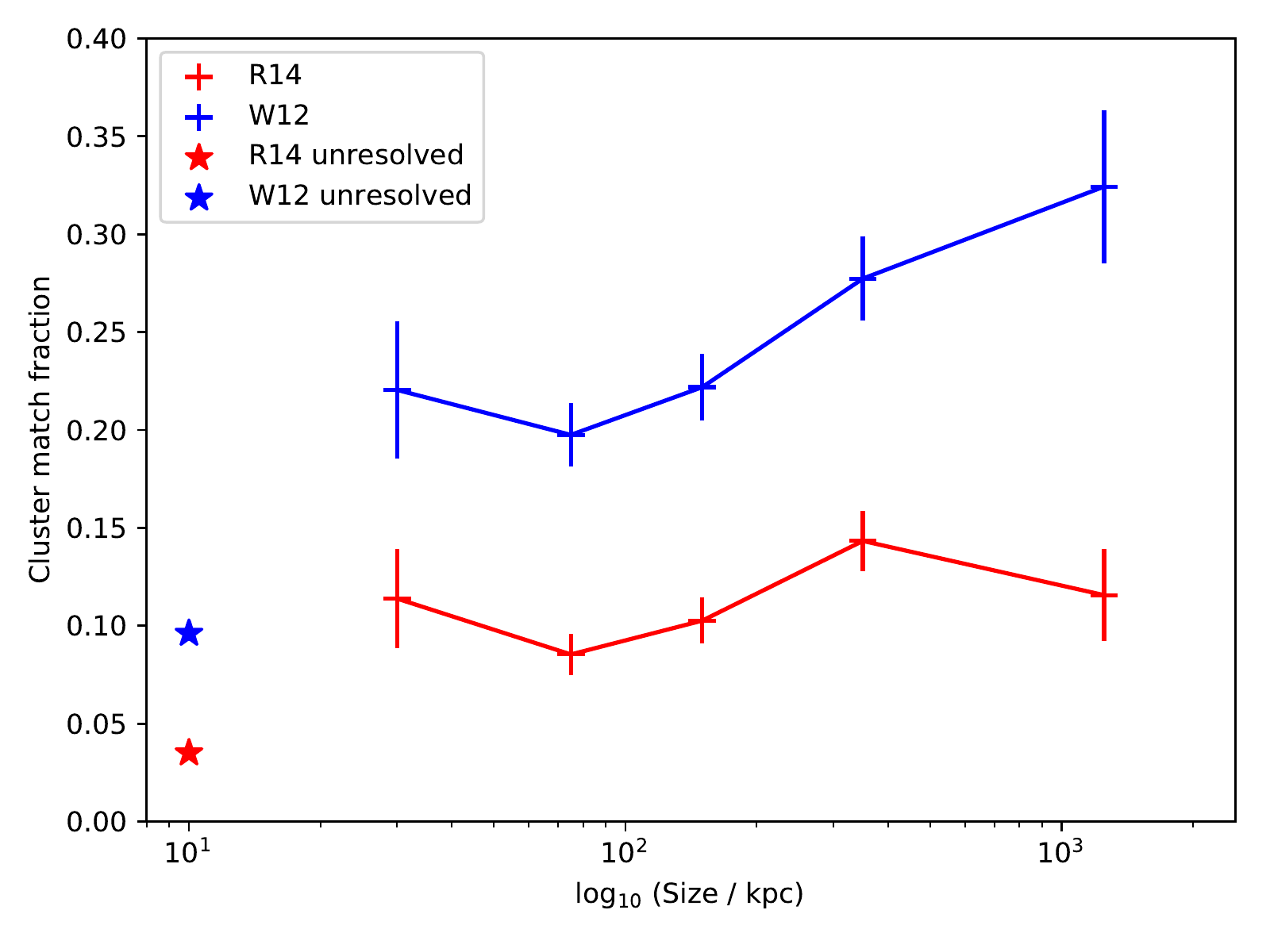}	
   \caption{Cluster association fraction (i.e. the fraction of AGN with a group/cluster match) for the AGN sample, as a function of 150-MHz luminosity (left) and source physical size (right). Red and blue indicate the R14 and W12 catalogues, respectively, and for the righthand plot stars indicate the mean association fraction for all unresolved sources.}
              \label{fig:agn_det}%
    \end{figure*}

\begin{figure*}
	\centering
      \includegraphics[width=8.8cm]{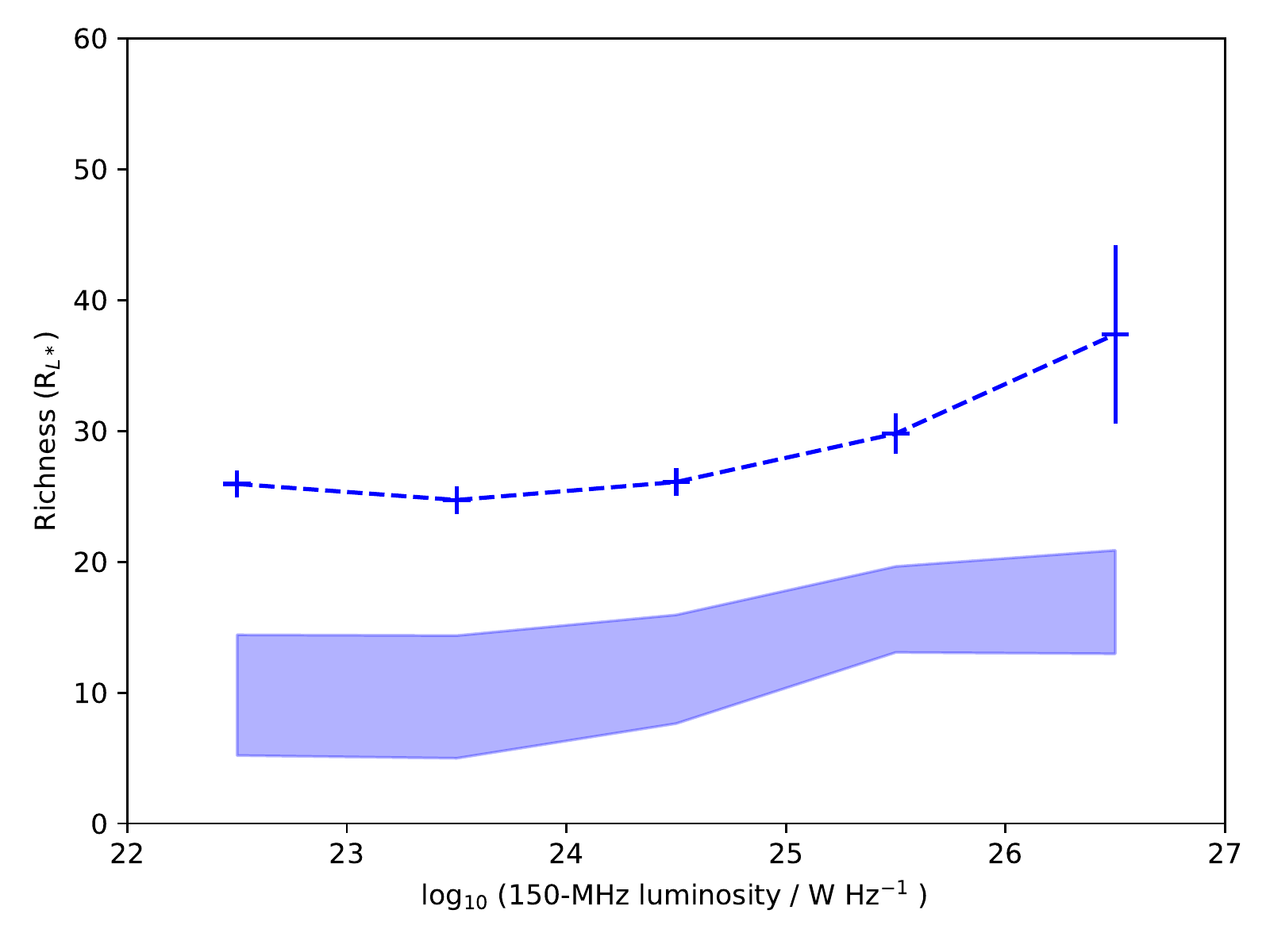}
   \includegraphics[width=8.8cm]{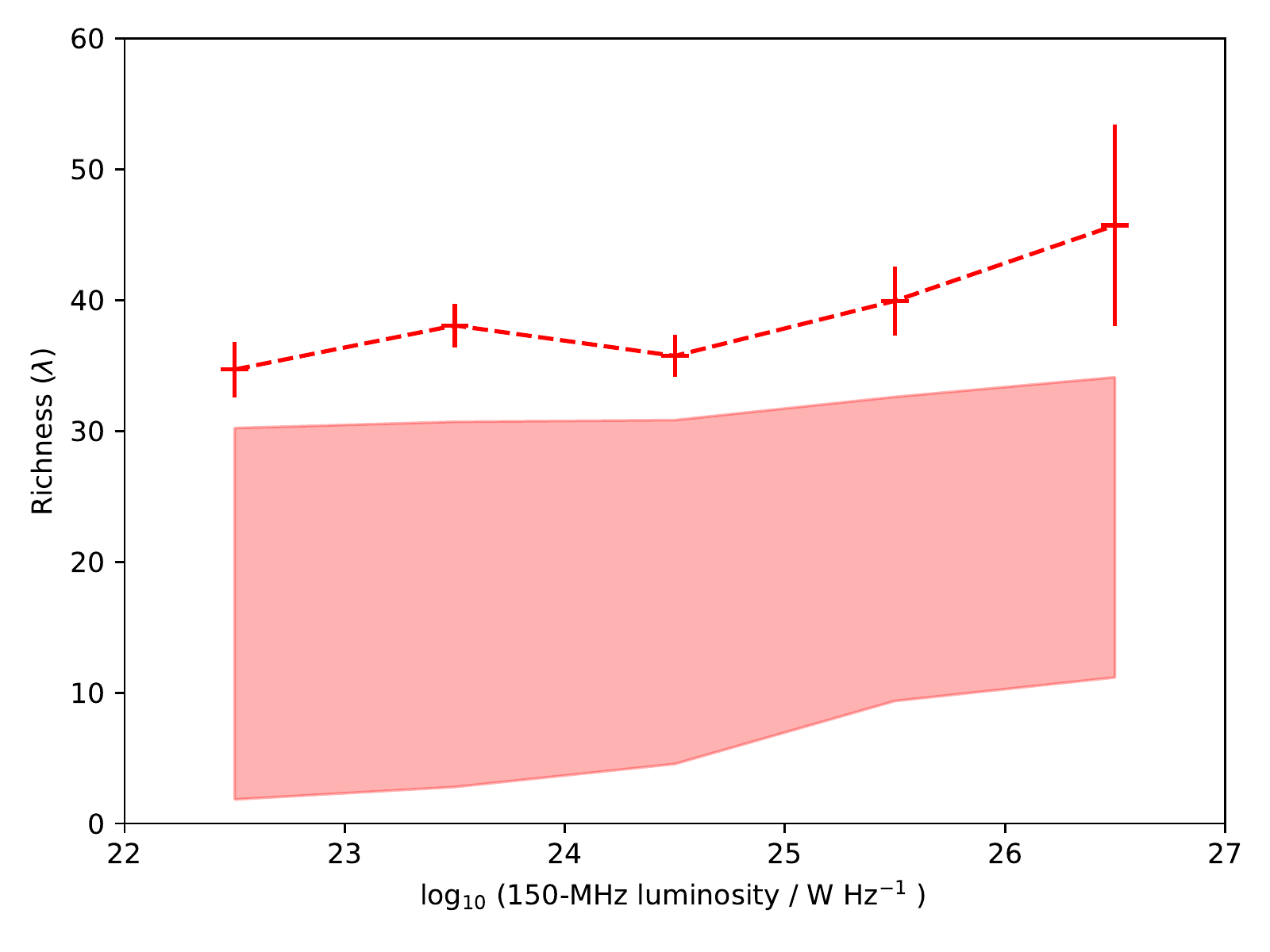}
	\includegraphics[width=8.8cm]{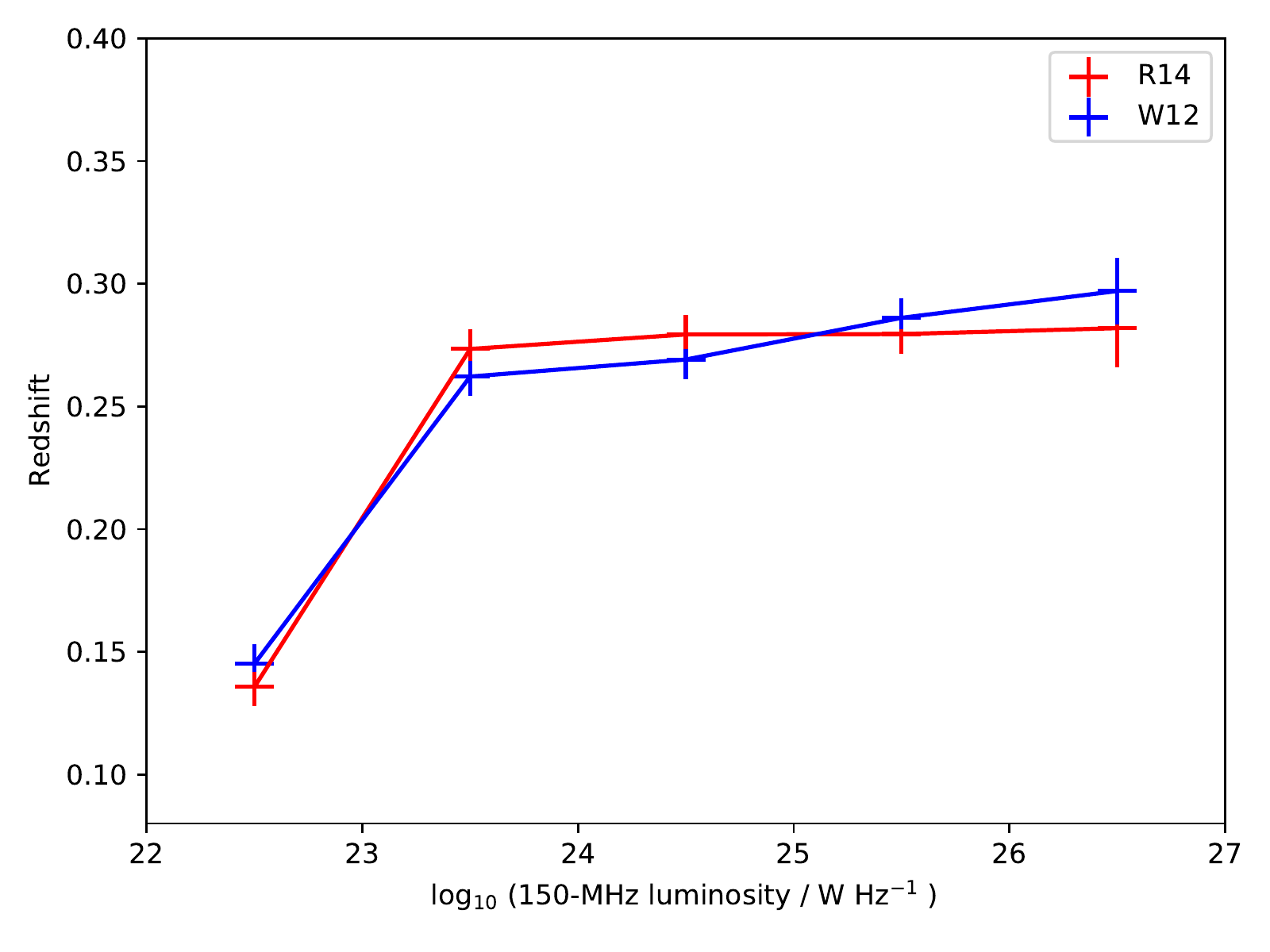}
    \includegraphics[width=8.8cm]{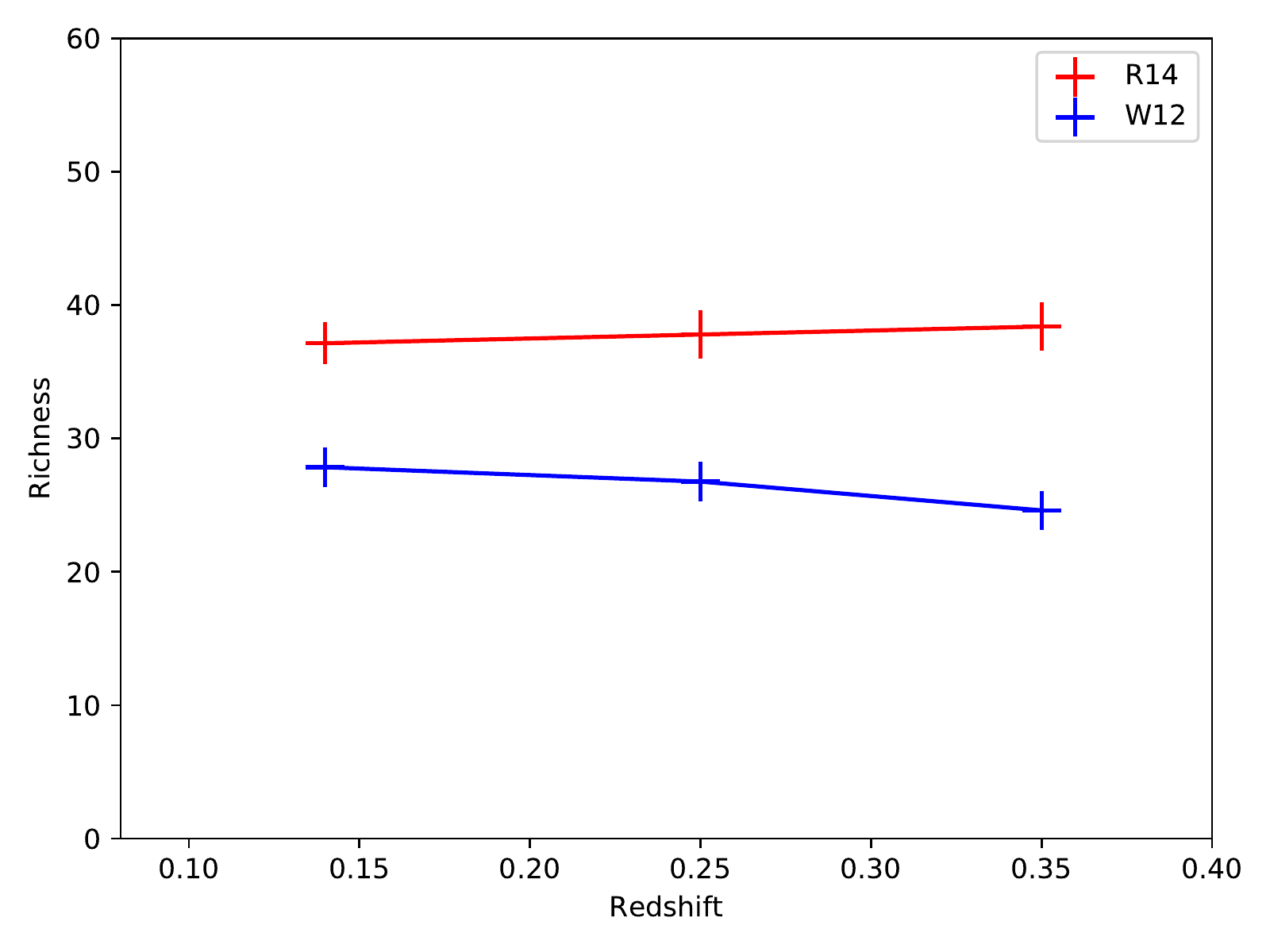}
    \includegraphics[width=8.8cm]{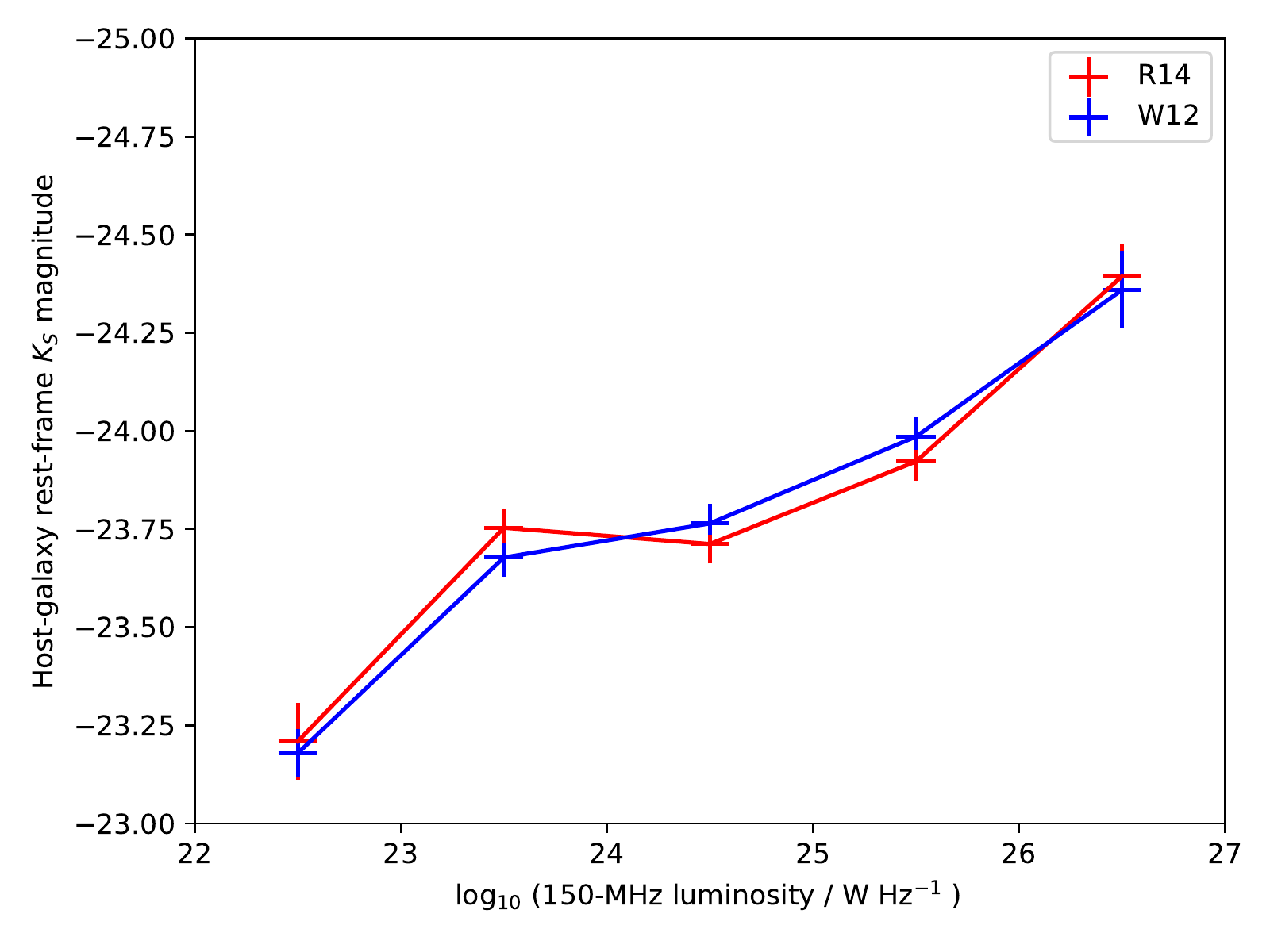}
   \includegraphics[width=8.8cm]{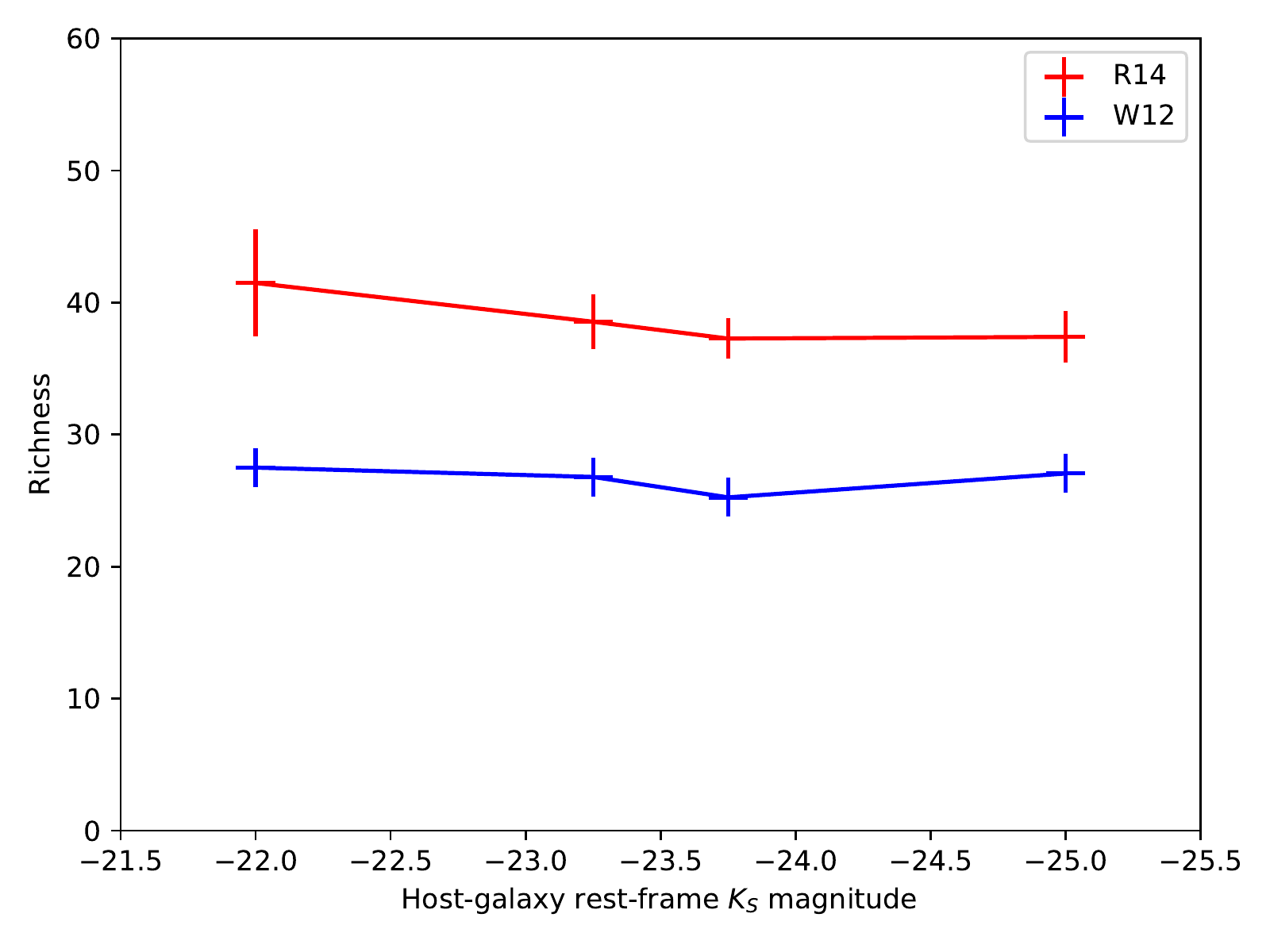}
     \caption{Relationships between environment richness and 150 MHz radio luminosity for the AGN with detected environments (top), between redshift and radio luminosity/environment richness (middle), and between host-galaxy rest-frame $K_{S}$ magnitude and radio luminosity/environment richness (bottom). Red and blue show results based on the R14 and W12 catalogues respectively. In the top panel, the shaded areas show the area of parameter space available when limits on the undetected environments are accounted for, with a lower bound given by the mean richness if AGN with an undetected environment are assumed to have an environment of richness zero, and an upper bound given by the mean richness if the richness for undetected to environments is assumed to be the richness detection limit.}
              \label{fig:agn_properties}% 
\end{figure*}

In total 1,089 AGN (12 percent) are found to be associated with W12 clusters with a probability $> 50$ percent, and 899 (10 percent) with a probability $> 80$ percent, while 456 AGN (5 percent) are found to be associated with R14 clusters with a probability $> 50$ percent, and 404 (4 percent) with a probability $> 80$ percent (we use a threshold of 80 percent for our high probability category so as to include the best photometric matches, for which the size of the errors in most cases makes it impossible to achieve a higher threshold of 90 or 95 percent). Hence, in total, $\sim 10$ percent of the LoTSS AGN have a high-probability group/cluster association. 

We now consider the relationships between the properties of LOFAR-DR1 AGN at $0.08<z<0.4$ and SDSS-identified galaxy groups and clusters. We first discuss the environments of the full AGN sample as a function of radio-source properties in Section~\ref{sec:envs} and then discuss the 150-MHz properties of the full cluster population(s) in Section~\ref{sec:clus}. Where we calculate mean properties, this corresponds to the arithmetic mean -- in the case of radio luminosity the mean is calculated in logarithmic space. Uncertainties on mean quantities are the 68 percent confidence intervals determined via bootstrapping. Trends reported as significant are robust to the choice of binning.

\subsection{The environmental properties of the LoTSS-DR1 AGN}
\label{sec:envs}
Fig.~\ref{fig:agn_det} shows the fraction of AGN with a group/cluster association for both cluster catalogues as a function of 150-MHz radio luminosity and as a function of source size, calculated by summing the probabilities of association over all potential matches. We find a strong trend for increasing group/cluster association fraction with radio luminosity. It is important to note, though, that even at the highest luminosities ($L_{\rm 150MHz}> 10^{26}$ W Hz$^{-1}$) more than 60 percent of the LoTSS AGN do not have a group/cluster association in either the W12 or R14 catalogues-- i.e. there is a substantial population of powerful radio galaxies in environments of halo mass $M_{200} < 10^{14}$ M$_{\odot}$ (a conservative estimate of the limiting richness for completneness of the W12 catalogue in the redshift range considered). 

We also find a clear relationship between size and detection fraction. The unresolved radio sources (indicated by star symbols in the right-hand panel of Fig.~\ref{fig:agn_det}) have a low cluster association fraction; for the resolved sources a trend of association fraction with size continues up to the largest sources for the W12 catalogue (departure from a straight line is significant at $>99$ percent level), but for the R14 catalogue there is no significant trend with size for the resolved sources. 

The trends in cluster association fraction with luminosity and size (Fig.~\ref{fig:agn_det}) are not independent -- the low association fractions for the unresolved sources are similar to the values for the lowest radio luminosity bins, and \citet{hardcastle18b} show that the LoTSS surface brightness limit means that large sources cannot be detected at these radio luminosities. The dramatic difference in group/cluster association fraction for the unresolved and resolved sources is interesting, and supports the conclusion that the majority RLAGN population of physically small ($<10$ kpc), low-luminosity ($L_{150}<10^{24}$ W Hz$^{-1}$) sources \citep[e.g.][]{sadler14,baldi15, hardcastle18b} are not young, or ``switched-off'' powerful radio galaxies, but it will be important to push to higher radio sensitivities to enable a better understanding of the low surface brightness population. 

\subsubsection{Trends with radio luminosity}

Having established that the cluster association fraction is related to radio luminosity, we next investigated whether the mean richness of the associated clusters is related to the AGN properties, as suggested by previous studies using smaller samples, different environmental measures, and/or narrower luminosity ranges \citep[e.g.][]{best04,ineson15,ching17}. Fig.~\ref{fig:agn_properties} (top) shows the mean richness as a function of radio luminosity for the two cluster catalogues, calculated as a weighted mean incorporating the association probabilities. The results for both catalogues hint at a trend for mean environmental richness to increase as radio luminosity increases -- for the W12 catalogue the trend is inconsistent with a straight line (i.e. uniform richness across all luminosities) at $>99$ percent condidence, but while the distribution for R14 is suggestive of similar behaviour there is no significant trend. Larger sample statistics as LoTSS expands to a wider sky area, and most importantly more sensitive environmental measures, are needed to confirm this trend. 

It is also important also to consider how these results could change if the large number of unassociated AGN in the sample are accounted for. We assumed an upper limit of $R_{L*}=12$ and of $\lambda=30$ for AGN with no association, corresponding to the richness at which the two surveys are complete over the redshift range of our sample. For objects with a cluster association we compared the richness limit with the product of the association probability and the richness of the potential association, and adopted the upper limit value where it was the higher of the two. The shaded regions in the top panel of Fig.~\ref{fig:agn_properties} show the possible parameter space in which the mean richness for the full AGN sample (both those with detected and undetected environments) could lie, bounded at the bottom by the mean richness if all unassociated AGN have an environmental richness of zero (a very conservative lower limit), and bounded at the top by the mean richness assuming all unassociated AGN have an environmental richness around the upper limit value. For the W12 sample a trend continues to be suggested although uniform richness with luminosity is not ruled out; for the R14 sample the range of possible parameter space is too large to draw conclusions.

We investigated whether this apparent trend could be driven by a mutual redshift dependence of the properties, considering only the matched AGN plotted in the top panel of Fig.~\ref{fig:agn_properties}. The middle panel illustrates that while there is a positive trend in the mean redshift of our luminosity bins (mainly across the first two bins, with a weaker trend continuing to higher luminosity for the \citet{wen2012} sample), there is no significant evolution of the mean richness of the sample with redshift (left middle panel). Hence this result is not driven by distance-dependence. We also repeated the analysis shown in Figs~\ref{fig:agn_det} and \ref{fig:agn_properties} including only objects with $L_{\rm 150MHz}> 10^{23.5}$ W Hz$^{-1}$, the luminosity at which sources are detectable at our upper redshift bound, and find no significant differences.

\subsubsection{The role of stellar mass}

A crucial question, then, is whether this trend is solely driven by the properties of the AGN host galaxies, or whether it requires additional links between AGN and large-scale environmental properties, such as via a role for fuelling from material originating in the ICM. It is known that the fraction of galaxies that are radio-loud is a strong function of stellar mass \citep[e.g.][]{best05}, and also that at fixed stellar mass there is an enhanced probability for a brightest cluster galaxy (BCG) to host a radio-loud AGN \citep{best07}. \citet{sabater18} have recently obtained the best constraints to date on the relation between radio-loudness and stellar mass, based on a sample selected from the same LoTSS-DR1 parent sample as this work. We do not have stellar masses for all of the objects in our sample, and so to investigate whether our trends between radio luminosity and richness may be driven by a mutual dependence of radio luminosity and richness on host-galaxy mass we used the rest-frame $K_{S}$-band magnitude (obtained for the LoTSS-DR1 sample as described by \citealt{duncan18}) as a proxy for stellar mass. 

We first considered whether there are trends in the mean host-galaxy (rest-frame) $K_{S}$ magnitude for the bins in radio luminosity used to generate the plots in the top row of Fig.~\ref{fig:agn_properties}, and find a strong trend in the mean host-galaxy magnitude per bin of radio luminosity (bottom-left panel of Fig.~\ref{fig:agn_properties}), consistent with the known relation of radio luminosity and stellar mass. Hence the more radio-luminous systems do have more massive hosts as well as richer environments than the lower luminosity AGN. However, if we consider the same population of AGN and examine the mean environmental richness in bins of host-galaxy magnitude (bottom-right panel of Fig.~\ref{fig:agn_properties}) we find that for this population of AGN there is no corresponding link between host-galaxy magnitude and environmental richness. In other words, for this LoTSS AGN population, more luminous host galaxies do not typically occupy richer environments. Hence the trend in AGN environment with radio luminosity is not driven by the effects of the host-galaxy mass.

\subsubsection{Trends with source size}

It might be expected that mean cluster richness also has some relation to AGN physical size. We investigated the mean richness for subsamples of different source sizes, but do not find any significant trend for the resolved sources (or a significant difference in the mean richness for associated systems only between the resolved and unresolved sources). If the W12 trend in cluster association fraction with size (Fig.~\ref{fig:agn_properties}, top right) does extend to the largest scales, so that a higher fraction of the giant radio galaxies are in rich environments than those on scales of a few hundred kpc, this is somewhat contrary to focused small-sample studies of giant radio galaxies that suggest they typically reside in poorer environments than the smaller objects \citep[e.g.]{machalski06,saripalli15,malarecki15}. Further work with deeper optical data (e.g PANSTARRS) and/or wider sky areas is needed to invesigate these relationships further.

\subsection{The radio-loud AGN populations of SDSS-selected clusters}
\label{sec:clus}

\begin{figure*}
\centering
	\includegraphics[width=8.5cm]{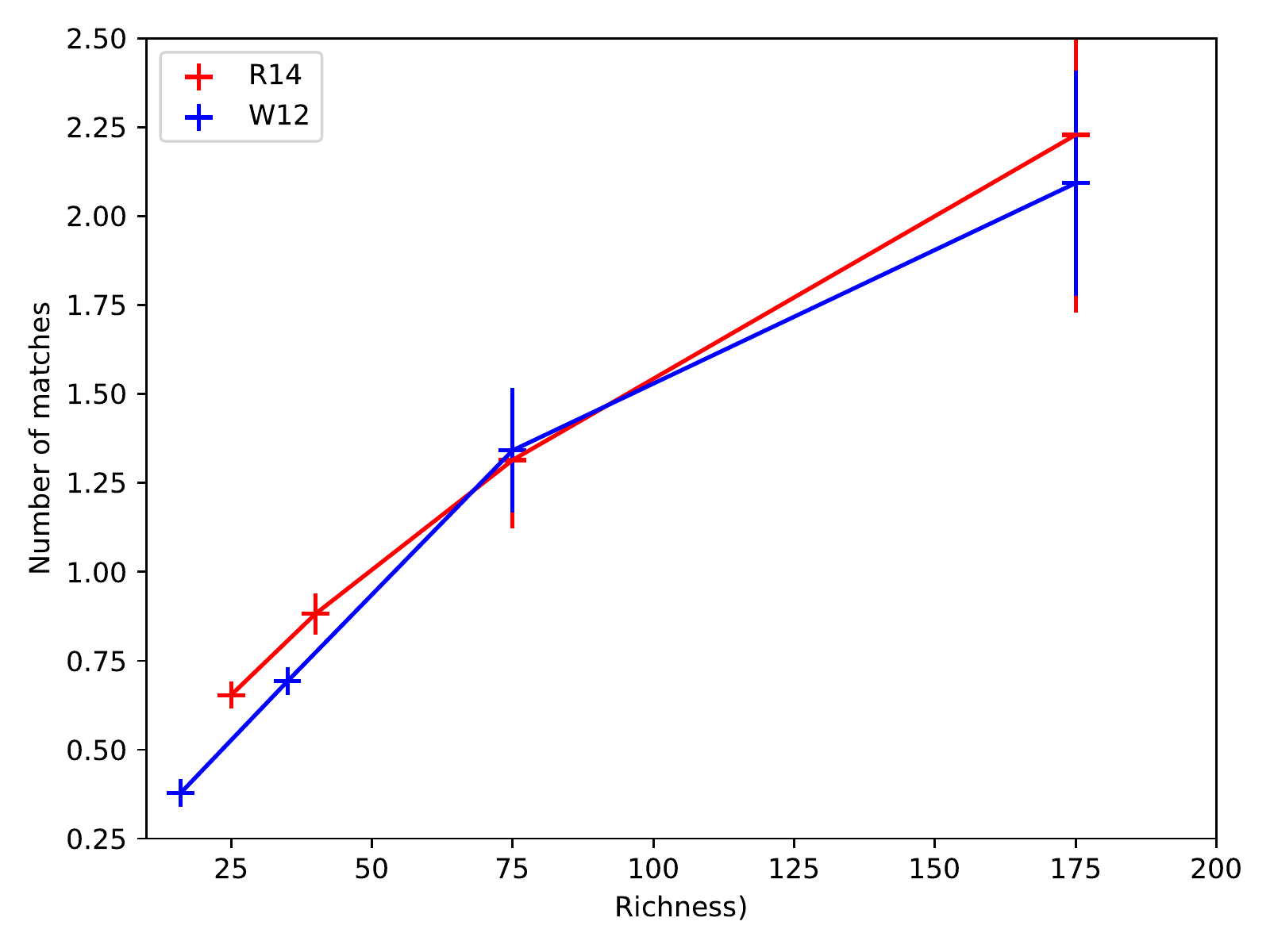}	
       \includegraphics[width=8.5cm]{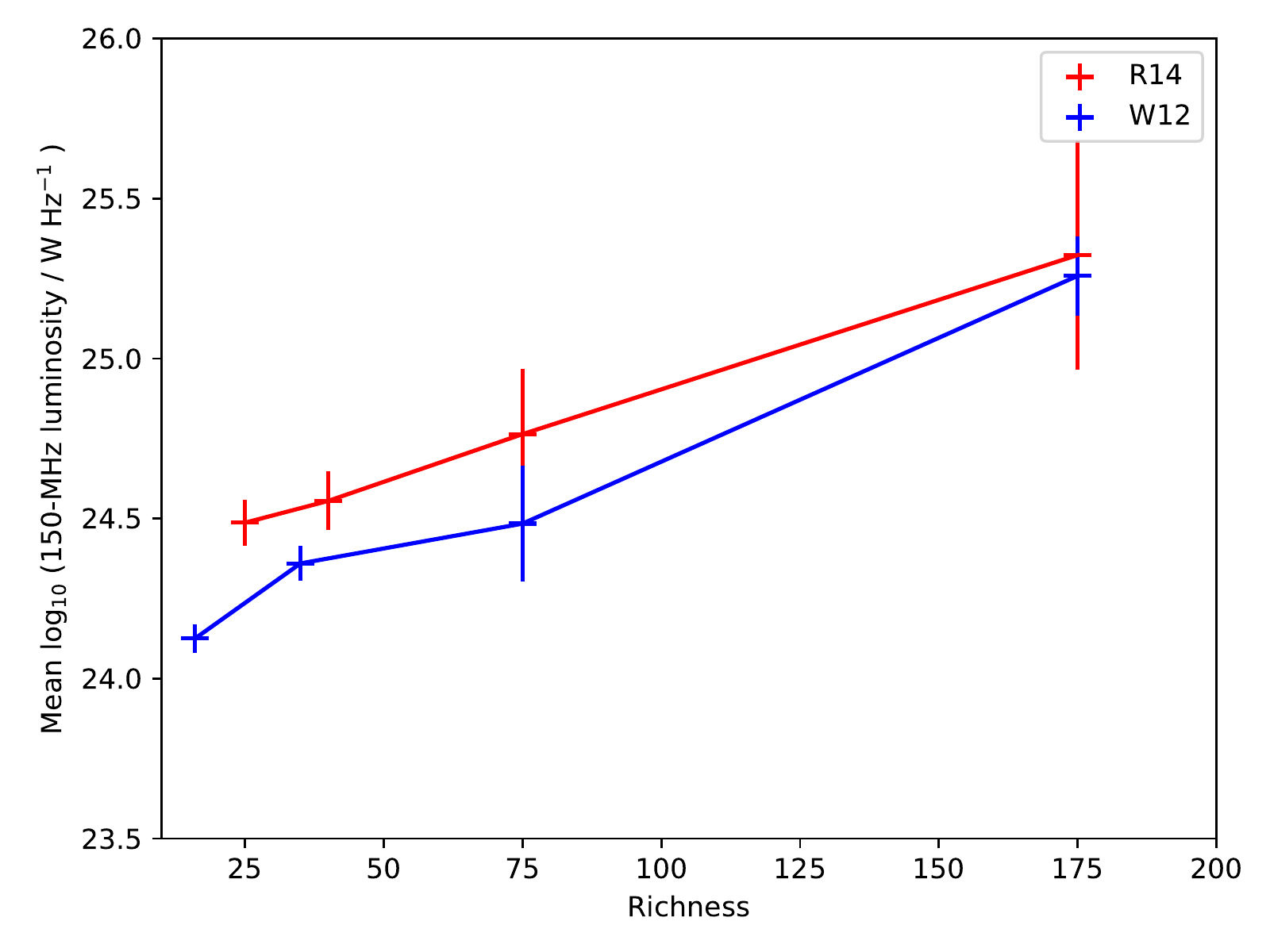}
   \includegraphics[width=8.5cm]{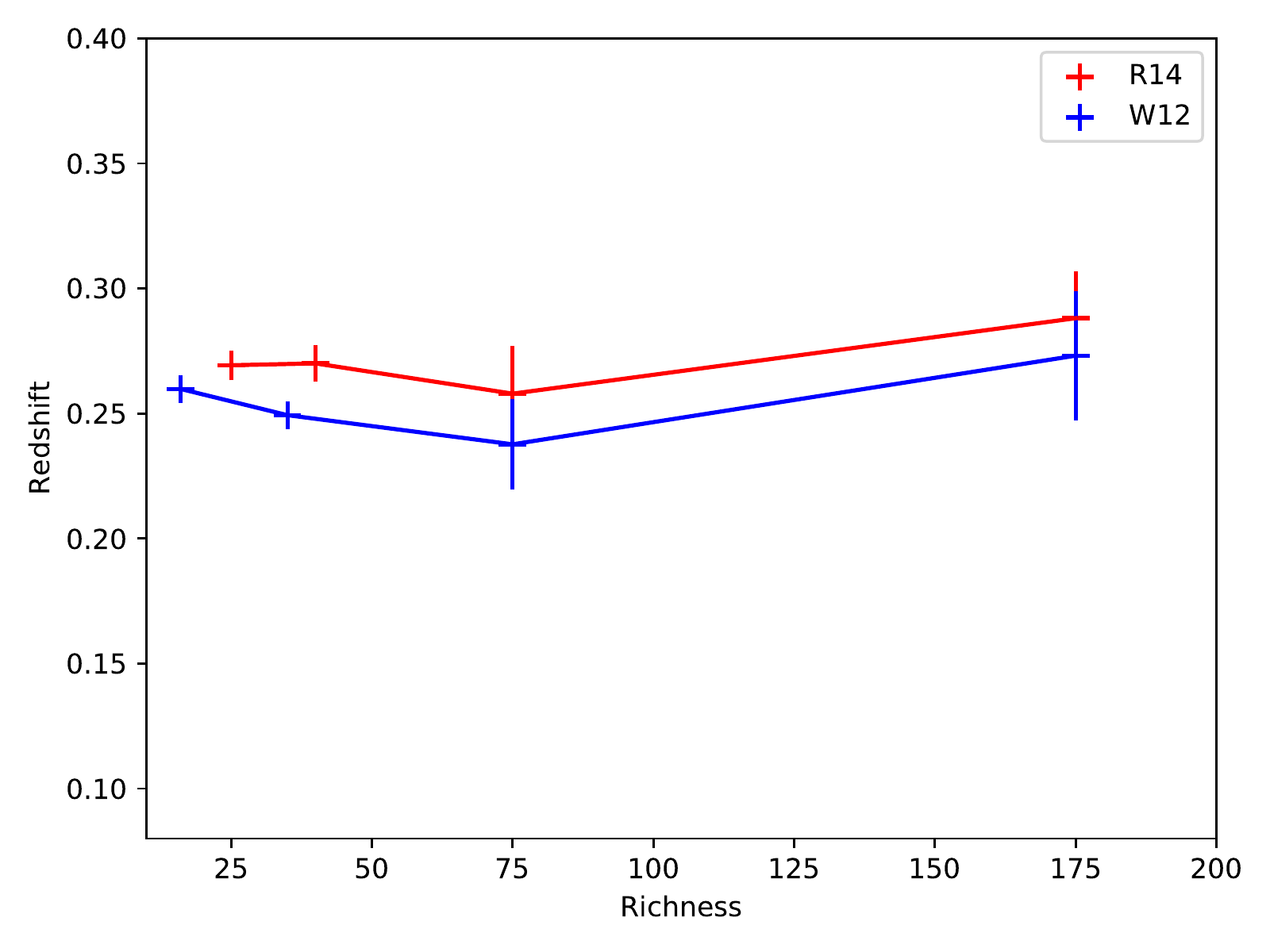}
      \includegraphics[width=8.5cm]{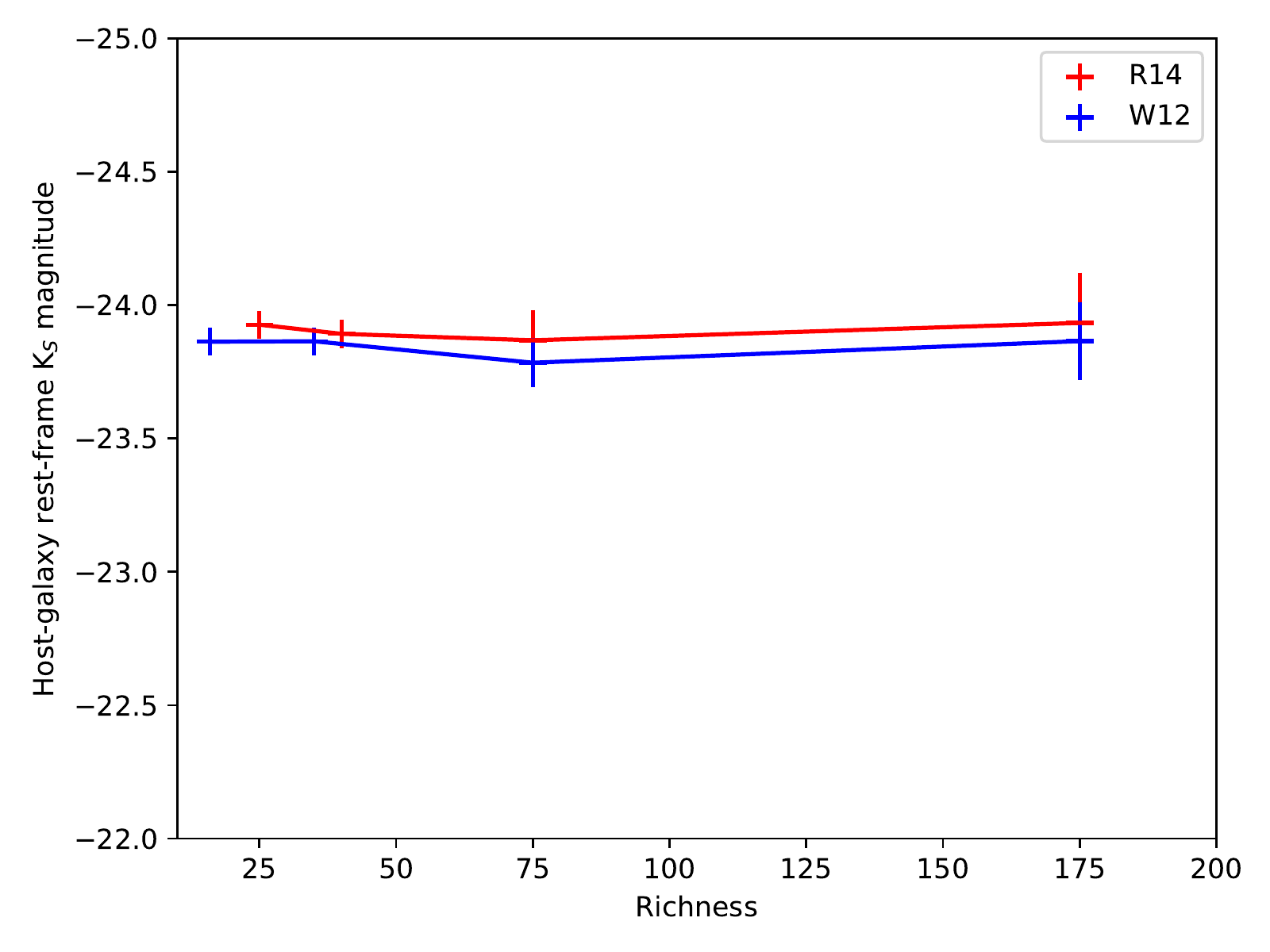}
      \includegraphics[width=8.5cm]{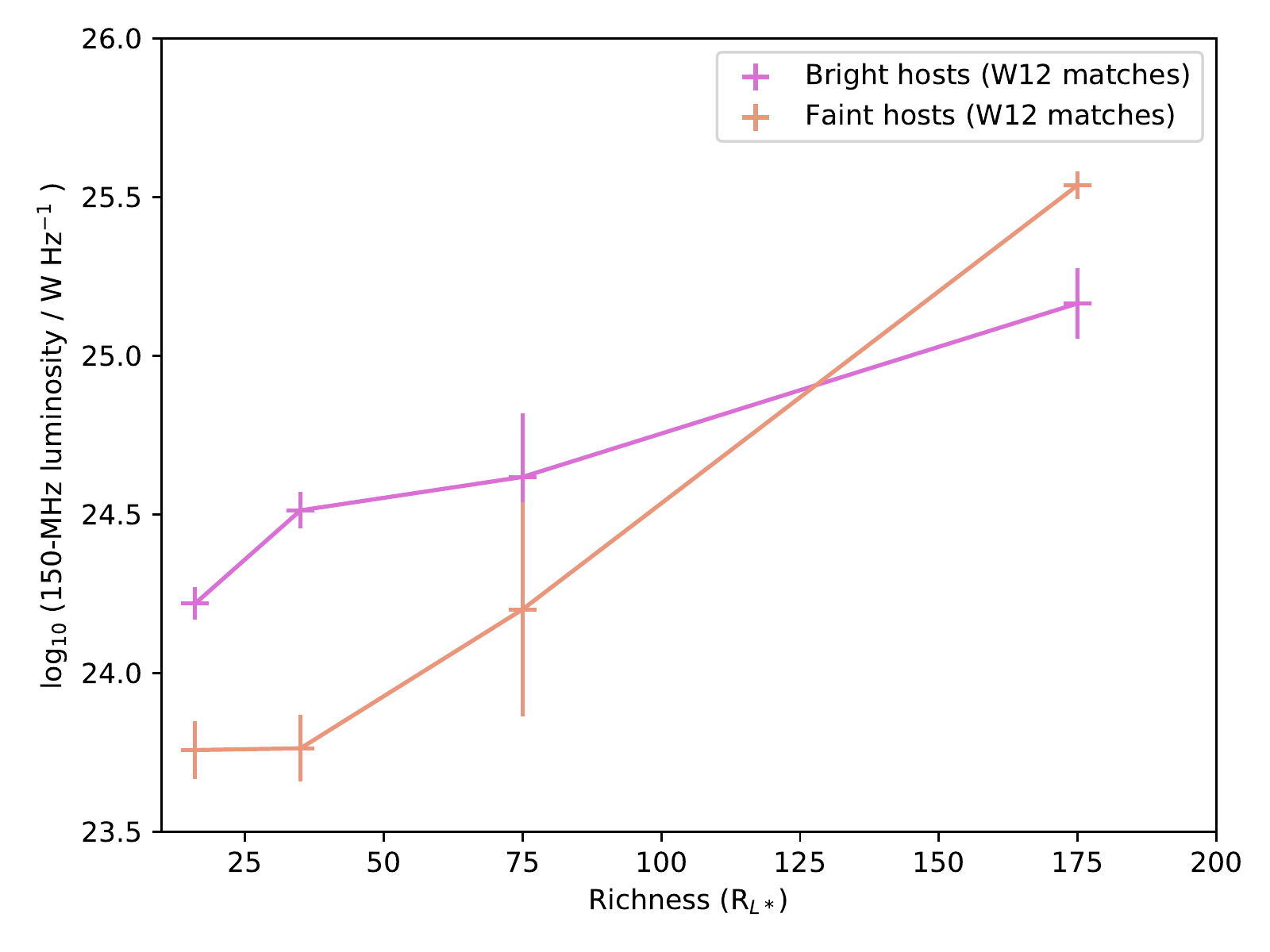}
      \includegraphics[width=8.5cm]{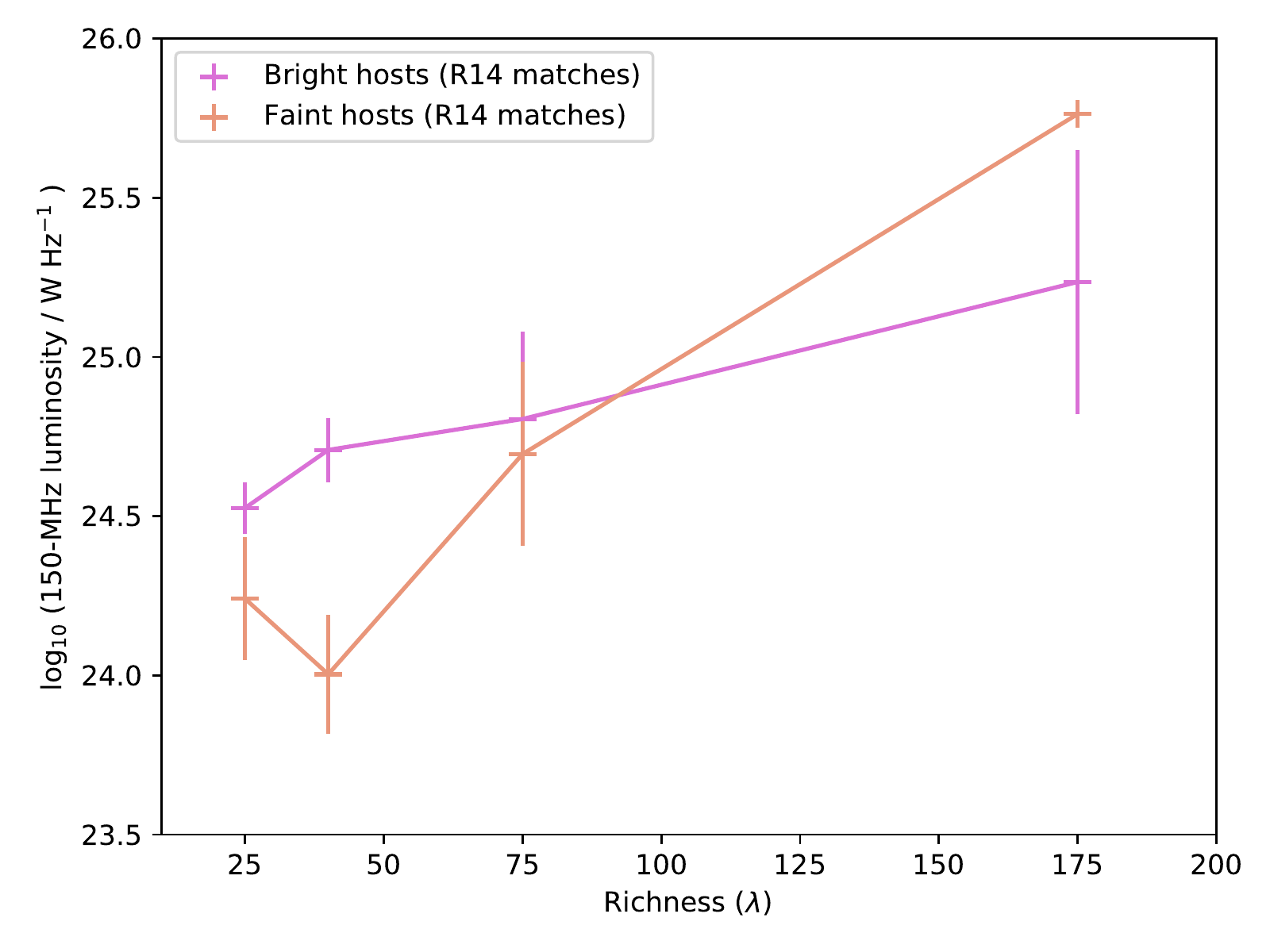}
   \caption{AGN properties of the the cluster population. Top: mean number of AGN matches per cluster (left) and mean 150-MHz radio luminosity of the brightest associated AGN (right) as a function of cluster richness. Middle panel: the relationship of richness with redshift (left) and host-galaxy $K_{S}$ magnitude (right) for the same richness bins as the top panel. Bottom: Mean radio luminosity versus richness as plotted in the top panel, but separated into two bins of host-galaxy magnitude (pink corresponding to $-26<K_{S}<-23.5$ and orange corresponding to $-23.5<K_{S}<-21.0$, where -23.5 is the median value.}
              \label{fig:clus_properties}%
    \end{figure*}

While the fraction of AGN having a high-probability group/cluster association is relatively small (3 and 10 percent for the two catalogues, respectively), the AGN association fraction for the cluster catalogues is high (consistent with previous studies). We find that 31 per cent of W12 clusters (733/2333 clusters in the $0.08 < z < 0.4$ range), and 47 per cent of R14 clusters (291/617 clusters in the $0.08 < z < 0.4$ range) have at least one high probability LOFAR-DR1 AGN match. The larger association fraction for the R14 catalogue, which has a higher richness threshold than W12, in itself points to a link between AGN behaviour and cluster richness.

To investigate further, we first considered whether the number of associated AGN for an individual cluster was related to cluster richness. Fig.~\ref{fig:clus_properties} (top left) shows that, for both cluster catalogues, there is a positive trend between mean number of associated AGN and cluster richness. If all galaxies were equally likely to host a radio-loud AGN then we would expect the number of AGN matches to increase as the number of cluster galaxies increases, as is observed here. If we interpret the richness estimator to be a rough indication of the number of potential AGN host galaxies, then for the poorest catalogued groups the probability of hosting an AGN is $\sim 3$ percent, while for the richest clusters it is $\sim 1$ percent. This apparent slight decrease in the probability that a typical cluster member galaxy hosts an AGN as cluster richness increases is likely to be driven by the comparatively high probability that a group/cluster central galaxy will host an AGN, thus providing a larger boost to the average probability for members of poorer systems.

We next considered the mean radio luminosity and mean source size of the brightest associated AGN and the largest associated AGN (usually, but not always, the same source) as a function of cluster richness. We find no significant relationship between source size and cluster richness, but a significant trend for richer clusters to host sources that are more radio-luminous, as shown in the bottom panel of Fig.~\ref{fig:clus_properties}. In the middle left panel of Fig.~\ref{fig:clus_properties} we show the mean redshift in the same richness bins, demonstrating that the richness bins do not have significantly different mean redshifts, and so the observed trends with radio luminosity cannot be driven by distance dependence. 

We again considered whether this relation could be a result of the known relation between host-galaxy magnitude and radio luminosity. The sample of AGN considered in the top right-hand panel of Fig.~\ref{fig:clus_properties} includes only the most radio-loud AGN in each cluster, and this population has a significantly different distribution of radio luminosity to the full AGN sample shown in the plots of Fig.~\ref{fig:agn_det} and \ref{fig:agn_properties}, with higher mean radio luminosity. We find that there is no systematic change in the mean host-galaxy magnitude with cluster richness for this population: the host galaxies of the brightest radio-loud AGN are typically comparatively bright ($K_{S} \sim - 24$). Hence the trend in radio luminosity of the brightest AGN with richness shown in the top-right panel cannot be driven by stellar mass of the host galaxy. As a further test we divided the sample into two bins of $K_{S}$ magnitude (bottom row of Fig.~\ref{fig:clus_properties}), finding that the trend is present for both the brighter and fainter half of the host galaxy population. 

Finally we also repeated the analysis shown in Fig~\ref{fig:clus_properties} including only objects with $L_{\rm 150MHz}> 10^{23.5}$ W Hz$^{-1}$, the luminosity at which sources are detectable at our upper redshift bound, and find no significant differences in our results.

\section{Discussion}
\label{sec:discussion}

We have examined the environments of the largest sample of local radio-loud AGN studied in this way to date, making use of recently compiled group and cluster catalogues with well-calibrated richness estimators. We have found several relationships that together suggest a complex interplay between radio luminosity, stellar mass and cluster richness, in line with previous studies. 

A key result is that the majority of the LoTSS ($0.08 < z < 0.4$) AGN inhabit large-scale environments poorer than $M < 10^{14}$ M$_{\sun}$, including more than 60 percent of the most luminous objects. This is not unexpected given previous studies indicating moderately rich groups are the preferred environment for radio galaxies, at least at low redshifts \citep[e.g.][]{best04,ineson15}, but highlights the importance of jet energy injection across a wide range of environments. We find evidence that the unresolved LoTSS AGN (with sizes typically below $\sim 20$ kpc for this $z$ range) occupy systematically poorer environments than the resolved radio galaxies (which have physical sizes from $\sim 20$ kpc to $> 1$ Mpc), and may well form a separate population \citep[e.g.][]{baldi15,sadler14,hardcastle18b}, although LoTSS is not sensitive to large sources at the lowest luminosities considered here. 

We have also shown that the probability of a cluster association, and the typical richness of the environment, where detected, are linked to radio luminosity. For the two SDSS cluster samples we have shown that number of associated LoTSS AGN, and the luminosity of the brightest associated AGN, increases with cluster richness. Despite the well-known strong association of AGN radio luminosity with stellar mass, we find that neither of these luminosity-richness trends are driven primarily by host-galaxy magnitude. These results provide strong indication that large-scale environment does play an important role in determining the observed AGN radio properties, at least for a subset of the population. 

It is essential, however, to emphasise that the cluster association fraction for the most luminous sources in LoTSS is less than 50 percent (Fig.~\ref{fig:agn_det}), and so there remain a large fraction of powerful radio galaxies residing in environments too poor to detect. Previous studies \citep[e.g.][]{gendre13,ineson15,ching17} suggest that many of these are likely to be primarily high-excitation radio galaxies, possibly having a different fuel source \citep[e.g.][]{heckman14,hardcastle18c} and therefore a much weaker coupling between radio properties and large-scale environment. For this sample we do not have reliable optical excitation (accretion mode) classifications, and so we do not attempt to address this question directly. However, the un-associated population of high luminosity radio galaxies is important: in some cases the amount of energy being injected by the radio galaxies may be sufficient to unbind a significant proportion of the halo gas \citep[e.g.][]{kraft07}. This potentially large population of systems likely to be over-compensating for the cooling of their hot-gas environments is not accounted for in cosmological simulations.

\subsection{The location of AGN in their large-scale environments}
\label{sec:location}
\begin{figure*}
   \centering
   \includegraphics[width=6.0cm]{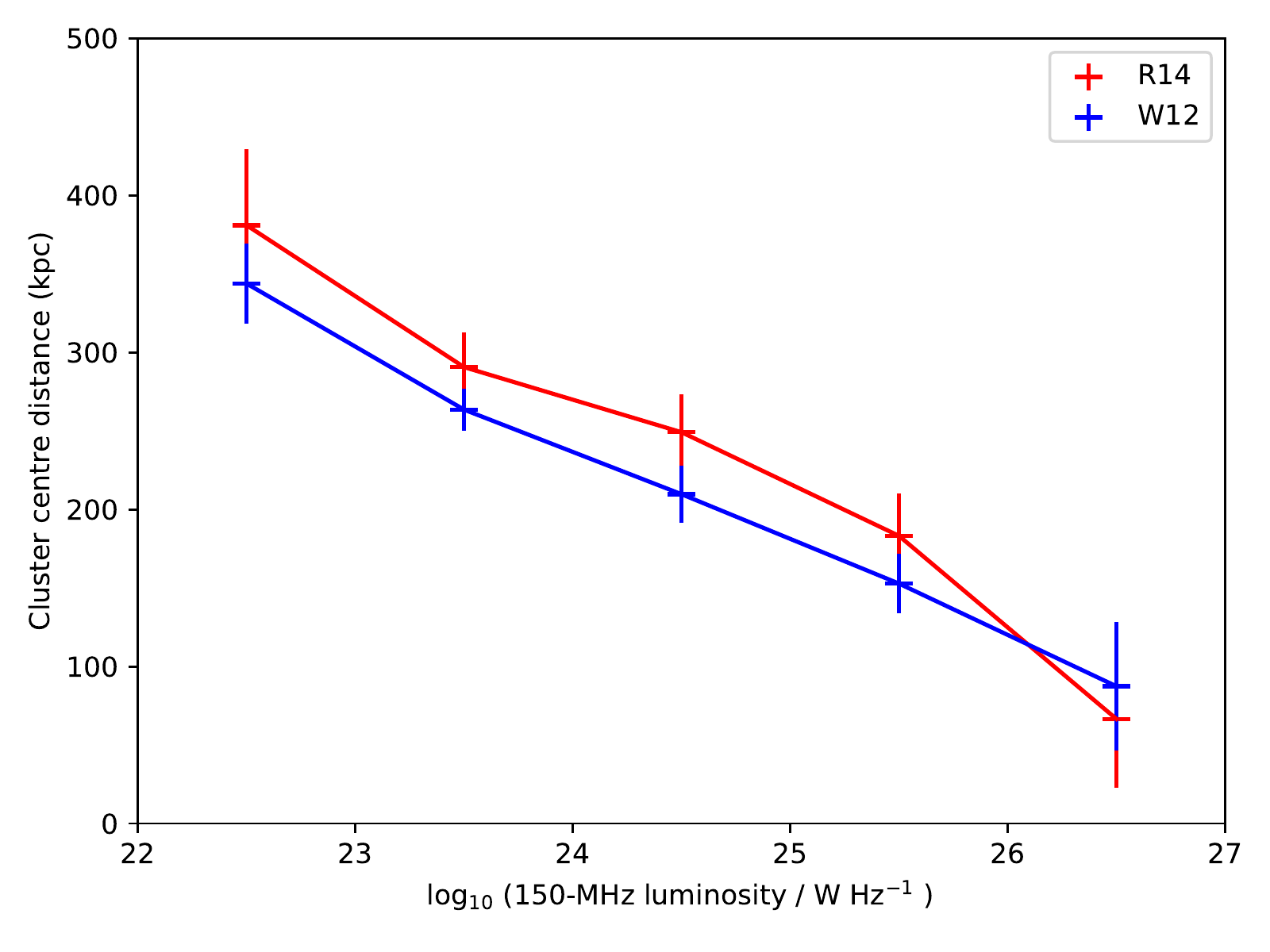}
   \includegraphics[width=6.0cm]{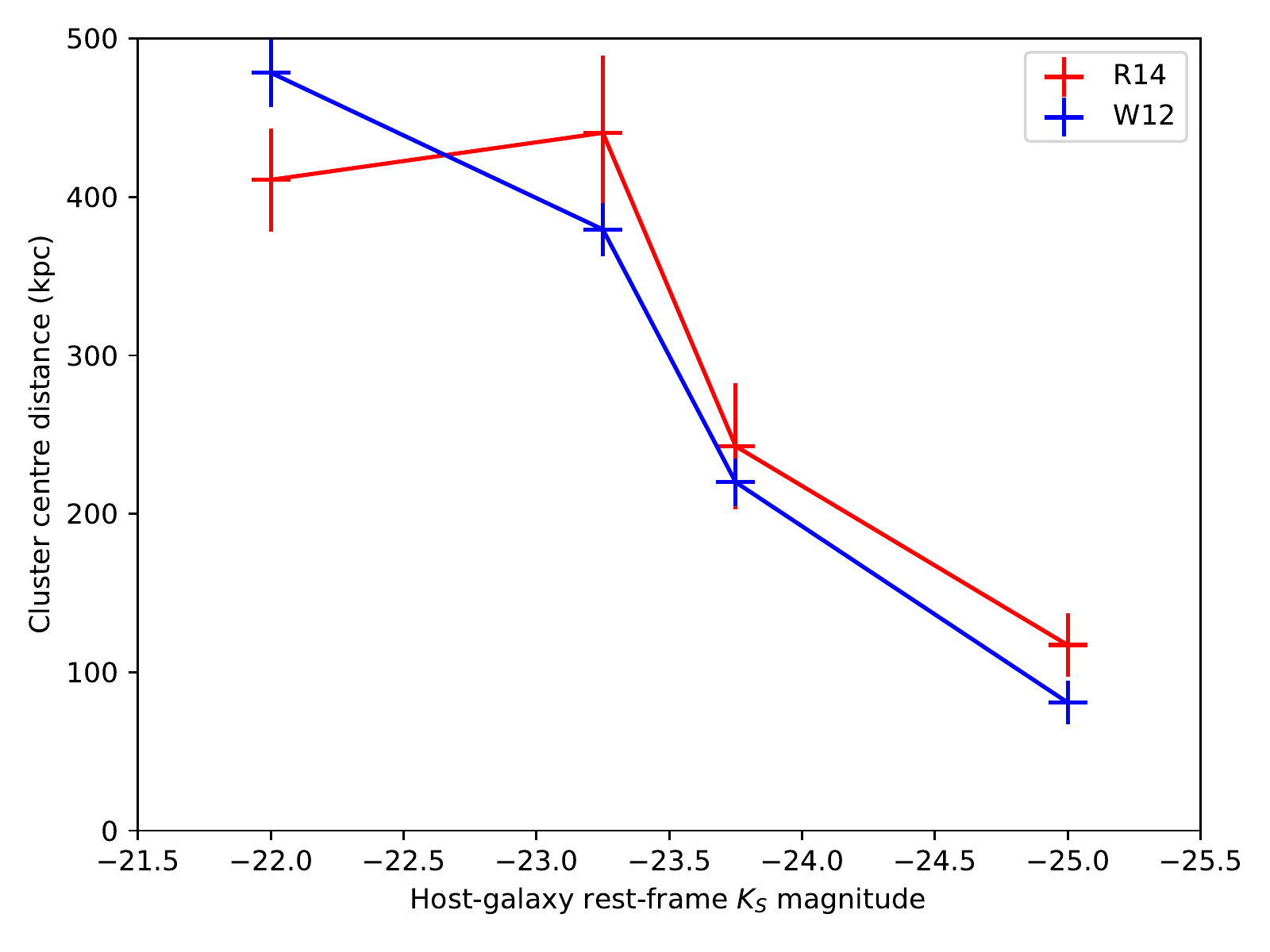}
   \includegraphics[width=6.0cm]{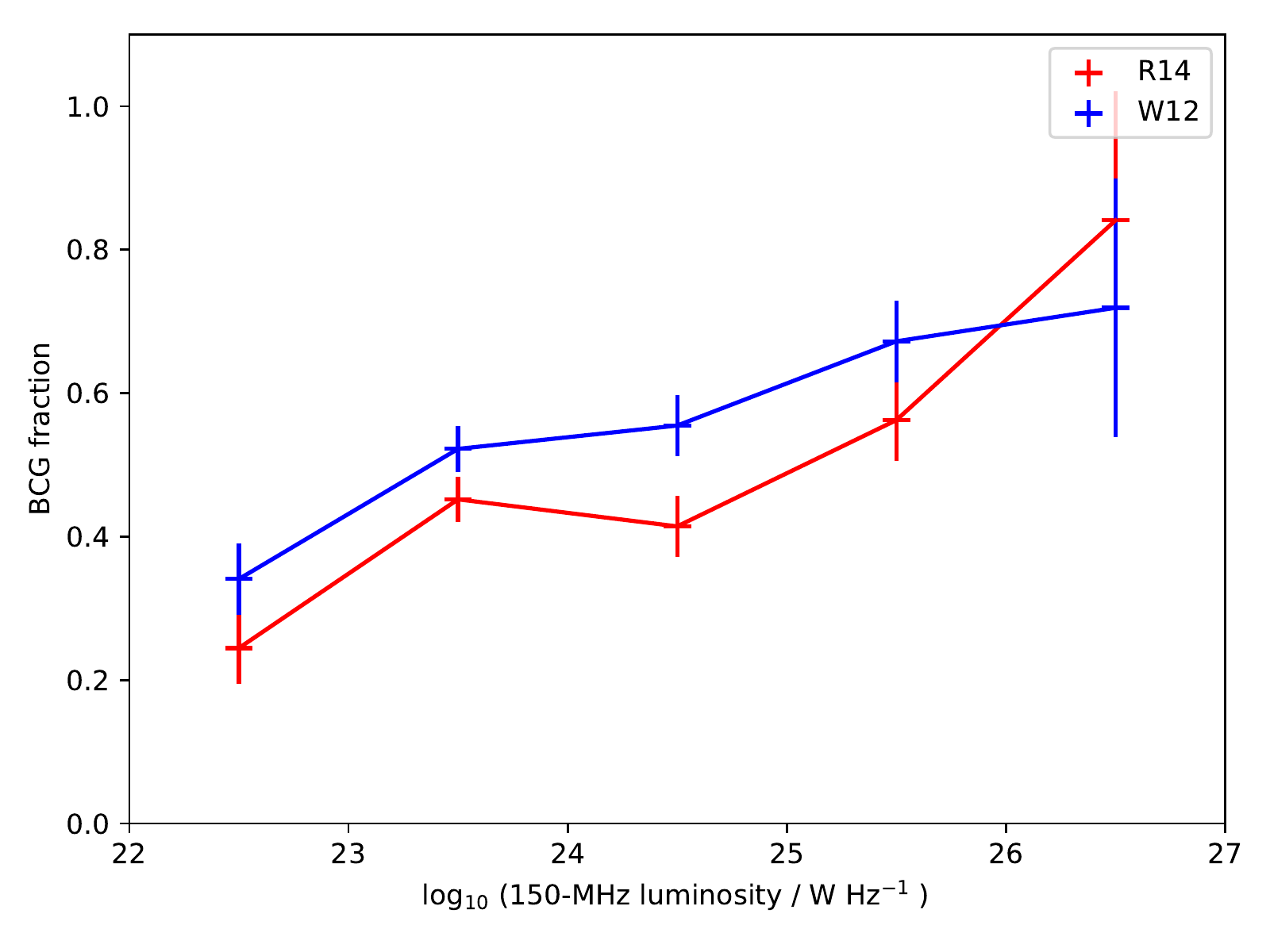}
   \caption{Left: Distance of matched AGN from the cluster centre, as a function of 150-MHz luminosity, for the R14 catalogue (red) and the W12 catalogue (blue); middle: as for the left-hand plot, but as a function of rest-frame $K_{S}$ magnitude; right: fraction of AGN associated with groups/clusters that are hosted by the cluster centre galaxy (BCG) as a function of radio luminosity (colours as for left panel).}
              \label{fig:agn_lum_sep}%
    \end{figure*}

A question of interest in the context of interpreting the observed luminosity/richness relationships is the connection between radio luminosity and source location within the group or cluster. Our approach of cross-matching with existing catalogues, rather than calculating overdensities around our AGN sample, enables us to investigate this. It is important to note, however, that while the AGN positions are well determined, the group/cluster centres are more uncertain \citep[see e.g.][for a discussion of the challenges of identifying the BCG for SDSS groups and clusters]{vonderlinden07}. Both \citet{rykoff14} and \citet{wen2012} report cluster coordinates corresponding to the brightest cluster galaxy (in the case of R14 the most probable central galaxy candidate) -- for the majority of groups and clusters this will be a very good estimate of the cluster centre, while for a small number the central galaxy may have been incorrectly selected, or offset from the centre of the mass distribution in a dynamically unrelaxed system. We therefore expect that the effects of incorrect cluster centering should be fairly minimal. In Fig.~\ref{fig:agn_lum_sep} (left) we show the mean cluster-centre separation for the AGN as a function of radio luminosity (here we consider all AGN with a group/cluster association -- i.e. the subsample considered in the analysis of Fig.~\ref{fig:agn_properties}). There is a strong trend for the least luminous radio-loud AGN to be found at significant distances, while the most luminous AGN are close to the centre. As shown in Fig.~\ref{fig:agn_properties} (bottom left), our radio luminosity bins do correspond to objects with systematically different host-galaxy masses. We therefore carried out a similar analysis of cluster-centre distance as a function of the AGN host galaxy luminosity, shown in the middle panel of Fig.~\ref{fig:agn_lum_sep}, which demonstrates that the trend shown in the left panel appears to be driven by host-galaxy properties. We note that including only objects with $L_{\rm 150MHz}> 10^{23.5}$ W Hz$^{-1}$, the luminosity at which sources are detectable across the full redshift range, does not alter the results significantly.

There is a high fraction of matched AGN in the sample whose separation is very small (Fig.~\ref{fig:cats}), typically where the AGN host and the galaxy defining the cluster centre are the same, with an identical spectroscopic redshift. The average separation in a given luminosity bin is likely to be at least partially driven by the fraction of AGN in that bin that are hosted by the cluster central galaxy. In the right-hand panel of Fig.~\ref{fig:agn_lum_sep}, we plot the fraction of all matched AGN (i.e. those associated with a cluster/group) that are hosted by the cluster-centre galaxy. As expected, the BCG fraction is significantly higher at high radio luminosities than at low luminosities; in other words, low-luminosity radio galaxies are less likely to be hosted by the central galaxy in their halo than those of high luminosity. We note that this result appears somewhat contrary to the recent findings of \citet{magliocchetti18}, but their analysis is for a smaller sample over a much larger redshift range. It is interesting to note that the BCG fraction across the luminosity range only changes by a factor $\sim 2 - 3$, while the mean separation distance changes by a factor $\sim 7$, and so a change in BCG fraction cannot fully explain the trend in the left-hand panel. This result, together with the, apparently host-galaxy independent, links between richness, cluster match fraction and radio luminosity (Fig.~\ref{fig:clus_properties}), suggests that location within the large-scale environment does play some role additional to that of stellar mass.

Finally, it is of interest to understand the population of radio-loud AGN in cluster outer regions in the context of cosmic ray and magnetic field injection into the ICM, and the seed population for radio relic emission. We therefore also investigated the relationship between source size and cluster location; however, we find no significant trend in the mean, with sources ranging from 20 kpc to Mpc scales found both in the centre and at large distances.

%Together with our conclusion from the previous section that there appears to be a relationship between radio luminosity and cluster richness {\it not driven by host-galaxy mass}, these results strongly favour a model in which {\it those powerful radio-loud AGN at $z<0.4$ that reside in groups and clusters} are being fuelled from material originating in the large-scale hot-gas halo -- i.e. irrespective of host-galaxy mass, the density of the surrounding ICM influences the power of the resulting jets (note that the environmental boosting effect described by \citealt{barthel96} cannot on its own account for these trends, as discussed by \citealt{hardcastle18a}). 

\subsection{Environment, radio morphology and luminosity}
\label{sec:morph}
A connection between radio morphology and environment has been suggested by many previous studies, and the interplay between jet power and environmental density is the favoured explanation for the FRI/II \citep{fanaroff74} morphological dichotomy \citep[e.g.][]{bicknell94,ledlow96}. We therefore made use of a morphological classification code that applies the standard FRI/FRII definition of centre vs. edge-brightening to LoTSS data \citep{mingo18} to obtain a subset of our $0.08 < z < 0.4$ AGN sample (656/8512, comprising 521 FRIs and 135 FRIIs) for which a reliable automated FR class could be determined (the remaining sources are either too small or too faint for a reliable automated morphological classification -- this is discussed further by \citet{mingo18}). We find that for W12, the FRI association fraction is 35 percent at $>50$ percent probability, and 30 percent at $>80$ percent probability, while the FRII association fraction is 23 percent at $>50$ percent probability and 20 percent at $>80$ percent probability. For R14, the fraction of FRIs having a group/cluster association is 17 percent with probability $>50$ percent, and 16 percent with probability $>80$ percent, while the fraction of FRIIs having an association is 10 percent with probability $>50$ percent, and the same at the higher probability threshold.  Hence, for both cluster catalogues the fraction of AGN with an association is lower for FRIIs than for FRIs. 

\begin{figure*}
   \centering
   \includegraphics[width=8.8cm]{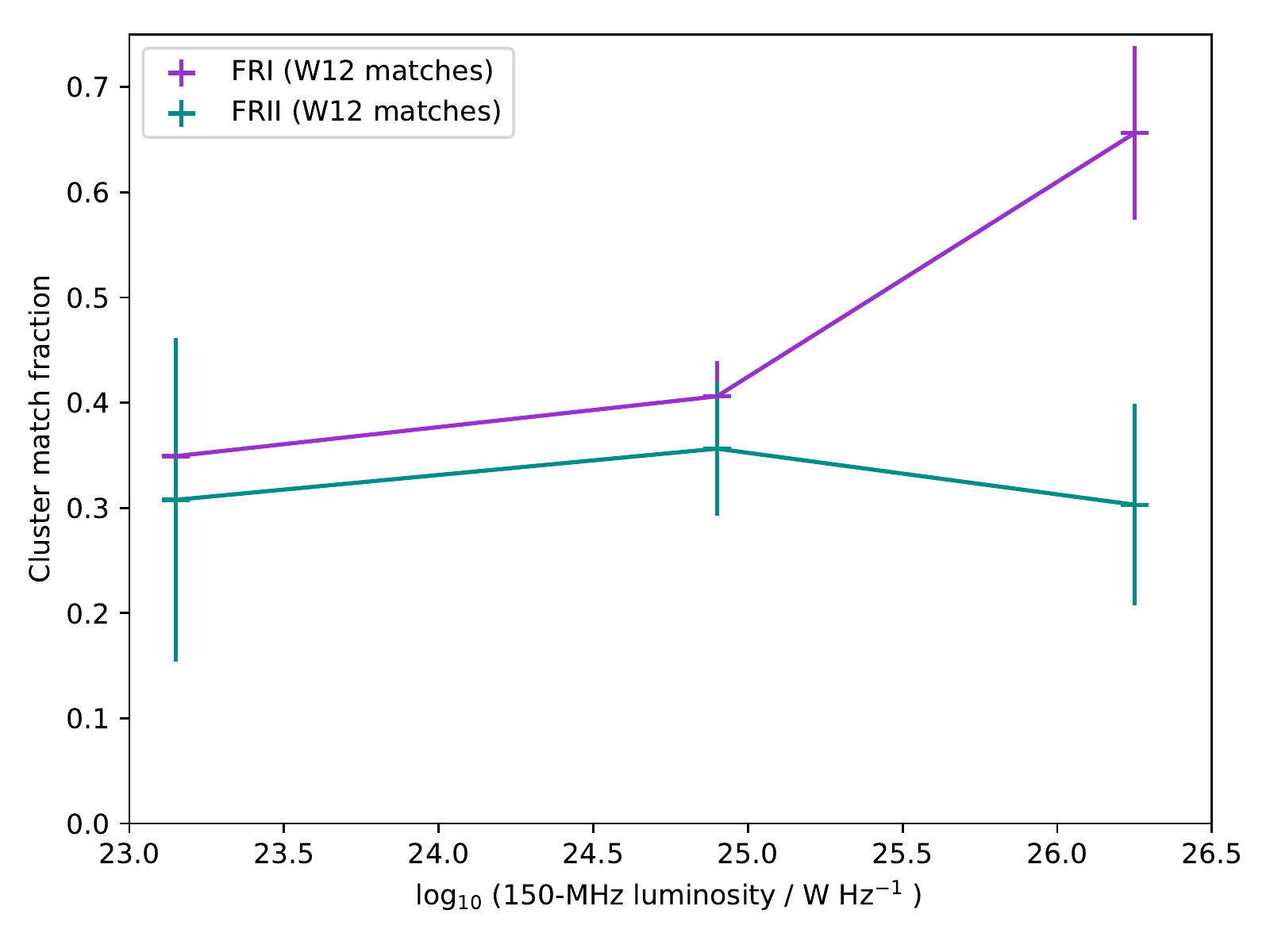}
	\includegraphics[width=8.8cm]{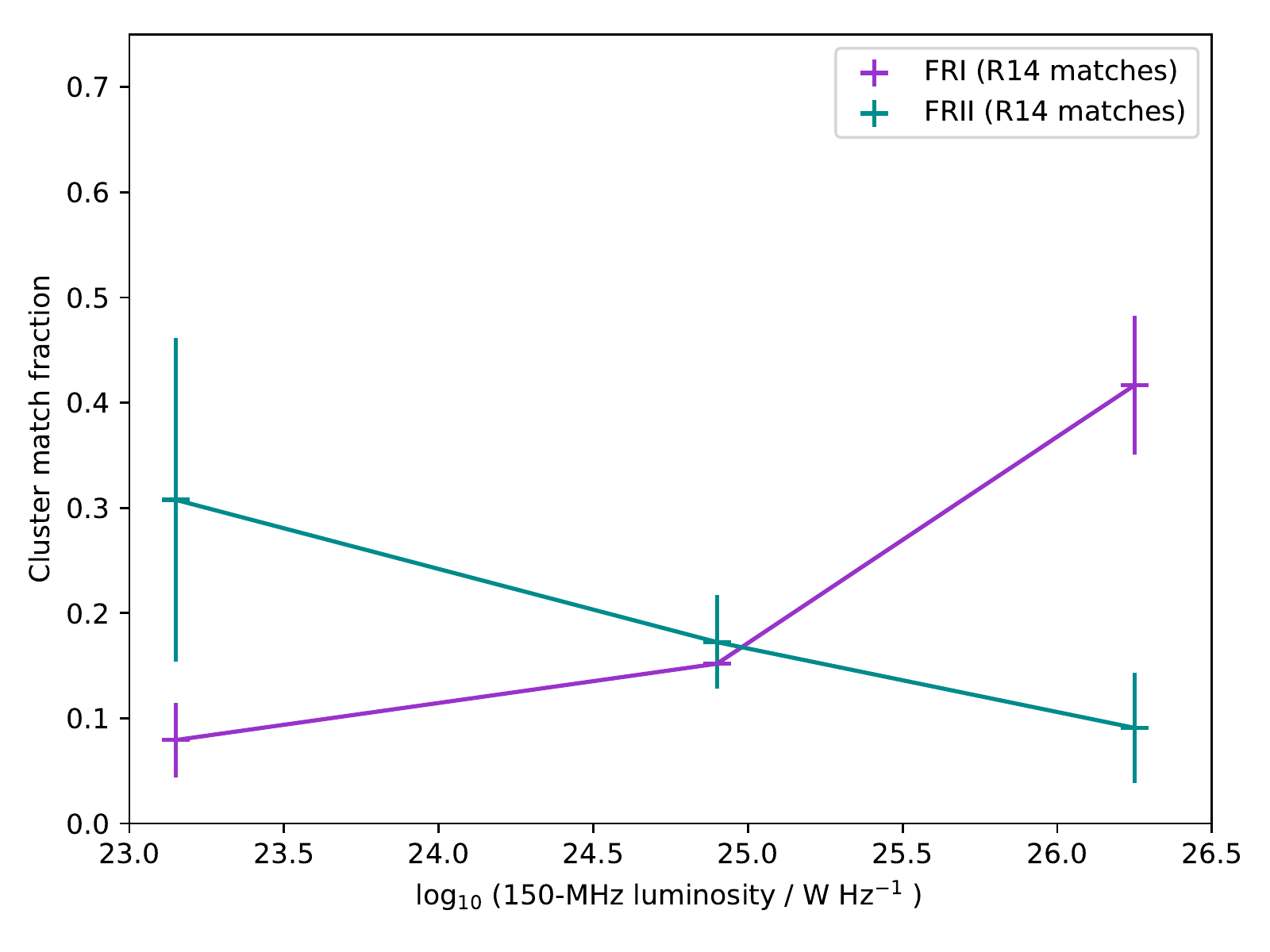}
   \includegraphics[width=8.8cm]{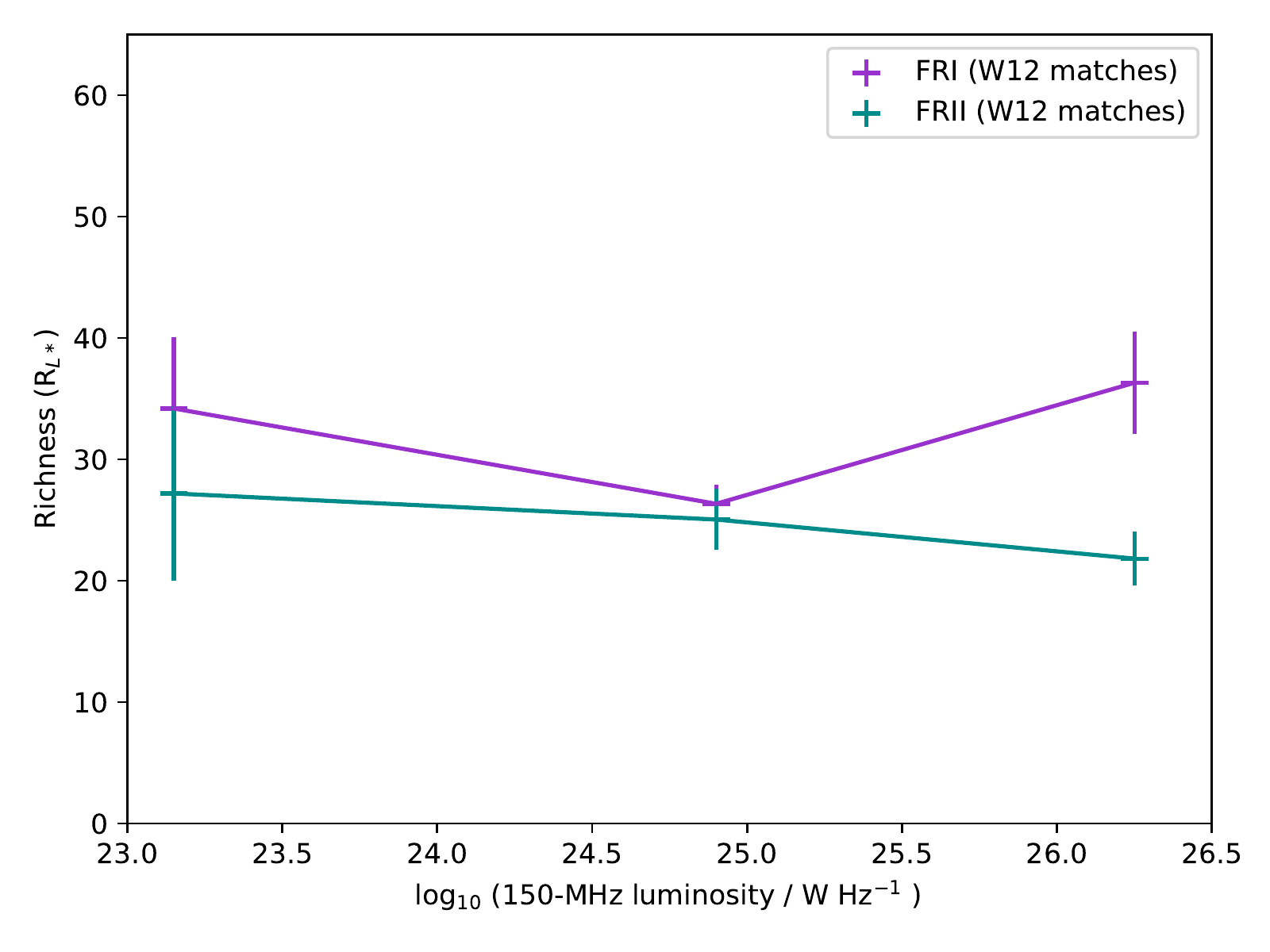}
   \includegraphics[width=8.8cm]{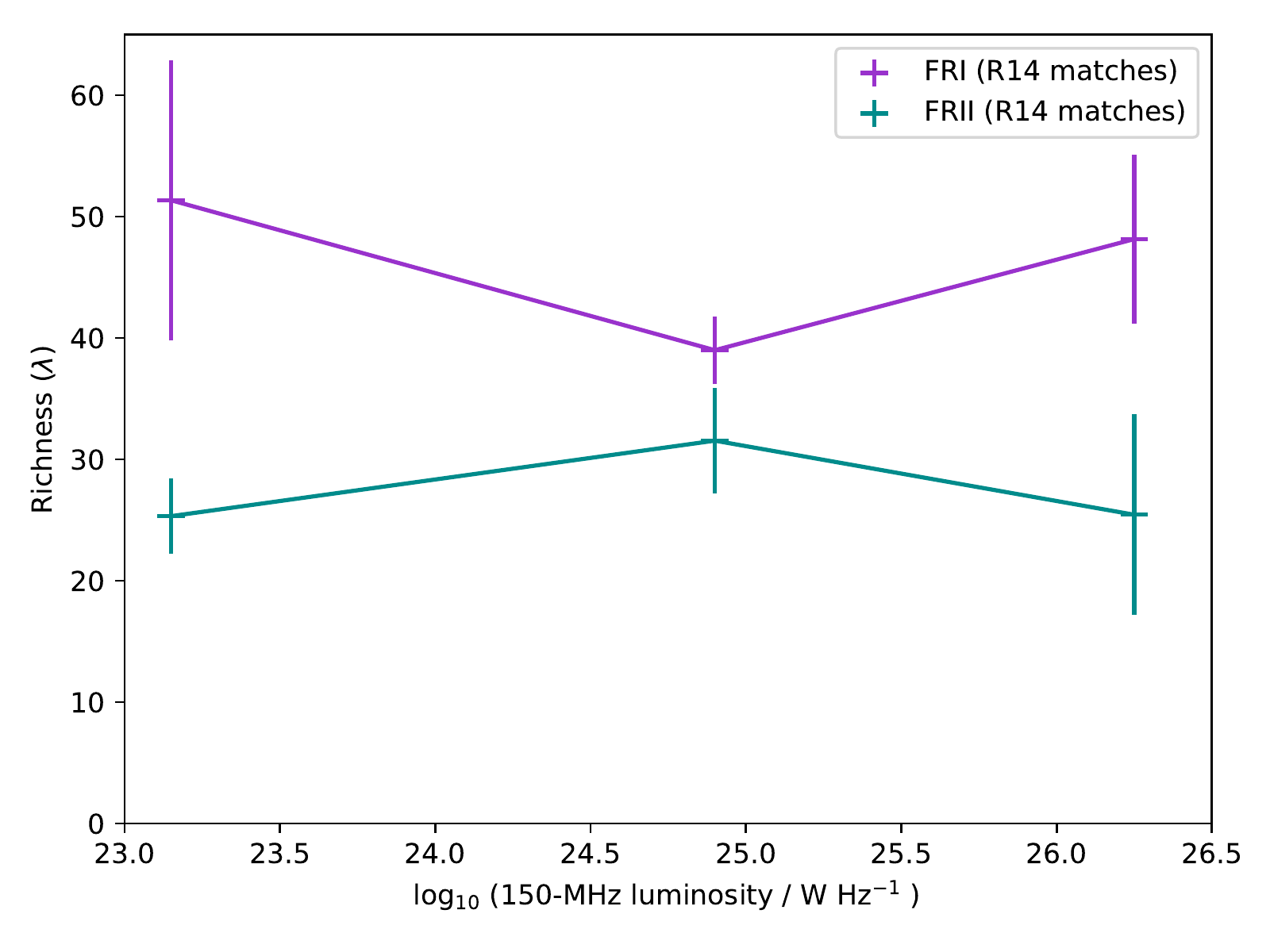}
   \caption{Top: Cluster association fraction as a function of radio luminosity for the FRI (magenta) and FRII (cyan) subsamples for the W12 (left) and R14 sample (right). Bottom: Mean richness as a function of radio luminosity for the FRI (magenta) and FRII (cyan) subsamples for the W12 (left) and R14 samples (right).}
              \label{fig:frclass}%
    \end{figure*}

In Fig~\ref{fig:frclass} we present the environmental detection fraction, and mean environmental richness as a function of radio luminosity separately for the FRI and FRII subclasses. The mean cluster association fraction (for both measures) increases with radio luminosity for the FRI subsample, while it is consistent with remaining constant for the FRII subsample. We note in particular that the cluster association fraction for luminous FRIs is substantially higher than for the luminous AGN population as a whole (Fig.~\ref{fig:agn_det}). We also investigated whether there is any difference in the typical cluster-centre distance, or relation between luminosity and cluster-centre distance, for the FRI and FRII subsamples; however, we found a large scatter in the mean cluster centre differences and so could not draw strong conclusions.

These results are consistent with previously found environmental differences between FRI and FRII radio galaxies. \citet{gendre13} found apparent environmental differences both for FRI and FRII sub-classes of AGN, and separately for low and high excitation sources (LERGs and HERGs, respectively), while \citet{lin10} find that high-excitation FRIIs are distinct in preferring poorer environments than both FRIs and low-excitation FRIIs. \citet{tasse08} also identified environmental differences linked to stellar mass and accretion mode. There are theoretical reasons to expect that effects related to both morphology and accretion mode are present: the favoured model for the FR dichotomy \citep[e.g.][]{bicknell94,ledlow96} predicts that, for a given jet power, the environmental richness controls whether the jet flow is disrupted to obtain an FRI morphology; while the prevailing view of LERGs as fuelled via material originating from the hot intracluster medium, in contrast to HERGs fuelled via a traditional accretion disk \citep[e.g.][]{hardcastle07,heckman14}, predicts an environmental difference related to accretion mode, independently of whether the jet evolves into an FRI or FRII morphology. As mentioned above, we do not have reliable accretion-mode classifications for our sample, but it is likely that the majority of the sources examined here -- i.e. those found to be associated with groups and clusters -- are low-excitation sources. For this reason, the environmental difference we find between FRIs and FRIIs at the high luminosity end of our sample may primarily originate from the effects of environment on jet evolution rather than being related to fuelling; however, we cannot firmly rule out the latter explanation. 

It is important to note that the most significant environmental difference is seen at the highest radio luminosities (where a significant difference is seen both in match fraction and mean richness for both samples), while the two catalogues give somewhat inconsistent results at lower luminosities. The highest luminosity bin encompasses the original Fanaroff and Riley \citep{fanaroff74} transition luminosity between FRI and FRII. As shown here, and discussed in detail in \citet{mingo18}, there is now considerable evidence for a population of lower luminosity sources with FRII morphology (as assessed at relatively low spatial resolution), for which the detailed jet dynamics and relation to the traditional -- more luminous -- classical double population is not clear. If it is only at higher radio luminosities that FRI and II sources inhabit systematically different environments, this could point towards the low-luminosity FRIIs being a somewhat different population to the well-studied more luminous FRIIs; however, further work is needed to investigate this question fully.

A final important point to consider in interpreting this apparent FRI/II environmental difference is that the jet power for FRI and FRII radio galaxies of the same radio luminosity is expected to be systematically different, because a larger fraction of the jet power in the FRIs is carried by non-radiating particles, as demonstrated by \citep{croston18}. Hence any intrinsic relation between large-scale environment and jet power would affect the radio appearance of the two sub-populations differently. An FRI radio galaxy with $L_{150} \sim 10^{25}$ W Hz$^{-1}$ could have a similar jet power to an FRII an order of magnitude more radio-luminous (or more). This effect could bring the environmental richness for FRIs and FRIIs of similar jet power into closer agreement, so that a single relationship between jet power and large-scale environment could be possible; however, previous work indicates that accretion mode is also a relevant, complicating factor \citep{ineson15,ching17}. A larger sample with sufficient numbers of more luminous FRIIs is needed to establish whether the FRI/II environmental difference we see here could be explained entirely in this way. 

\subsection{Caveats on the richness estimates and future plans}
\label{sec:richcav}
The optical cluster richness measures used in this work are relatively high-scatter proxies for cluster mass, as suggested by the large scatter between the two quantities (Fig.~\ref{fig:cats}). Although both \citet{rykoff14} and \citet{wen2012} calibrate their richness estimates against X-ray luminosity, temperature and/or SZ measures, the samples used for these calibrations are predominantly X-ray bright/relaxed clusters, and so the true scatter across the whole population may be underestimated. The large scatter in these measures mean that they do not provide accurate environmental richness measures for individual objects, but by deriving mean sample properties in bins of radio luminosity, source size and cluster richness we have been able to detect overall trends. A full exploration of the environments of the LoTSS DR1 AGN, and in particular the derivation of well-constrained environmental properties for individual objects would require a different approach. A crucial missing piece of information at present is a reliable estimate of the true scatter in halo mass for a particular range in radio luminosity -- given the large scatter in the richness/mass relation for the optical cluster catalogues used here, and the caveats about their calibration above, we do not attempt to do this here. In the future we intend to use the deeper PanSTARRS and WISE data, spectroscopic follow up of higher redshift samples and deep X-ray and optical/NIR data available for other LoTSS fields to obtain more well-constrained environmental properties for individual objects and large samples. 

\section{Conclusions}
\label{sec:conclusions}
%                                     Two column figure (place early!)
%______________________________________________ Gamma_1 (lg rho, lg e)
   
We have examined the large-scale environments of 8,745 radio-loud AGN selected from the LoTSS DR1 catalogue by cross-matching with SDSS group/cluster catalogues. This is the largest radio galaxy sample for which such an analysis has been carried out to date, and confirms previously suggested relationships between radio luminosity and environmental richness, obtained from studies based on smaller samples, narrower redshift ranges and different methods for estimating environmental richness. Specifically, we find that:
\begin{itemize}
\item 10 per cent of our AGN sample are associated (with high probability) with an SDSS-catalogued group or cluster.
\item The fraction of AGN with a group/cluster association increases with 150-MHz radio luminosity from $\sim 10$ per cent at $L_{\rm 150MHz} \sim 10^{22.5}$ W Hz$^{-1}$ to $\sim 30$ per cent at $L_{\rm 150MHz} \sim 10^{26}$ W Hz$^{-1}$.
\item More than 60 percent of even the most luminous radio galaxies in our sample do not have a group/cluster association -- i.e. there exists a substantial population of powerful radio galaxies residing in haloes with $M_{200} < 10^{14}$ M$_{\odot}$.
\item The mean cluster richness increases with 150-MHz radio luminosity -- AGN with $L_{\rm 150MHz} > 10^{25}$ W Hz$^{-1}$ are likely to be found in rich group/poor cluster environments, consistent with previous work \citep[e.g.][]{best04,croston08,ineson15}.
\item When the full group/cluster population is considered, the number of associated radio-loud AGN has a strong dependence on cluster richness, with the poor systems having a mean of $\sim 0.5$ associated AGN, while rich clusters have a mean of 1.5 -- 2 associated AGN; this trend is weaker than would be expected if all group/cluster members were equally likely to host an AGN (member galaxies in rich clusters are a factor 2 -- 3 less likely to host a LoTSS AGN than those in poorer systems).
\item The (logarithmic) mean radio luminosity and mean physical size of the brightest associated AGN also increases with cluster richness -- rich clusters are more likely to host an AGN with $L_{\rm 150MHz} > 10^{25}$ W Hz$^{-1}$ than poor groups, and the richest clusters are more likely to host large AGN.
\item For our associated AGN we find no relationship between AGN host galaxy rest-frame $K_{S}$-band magnitude and cluster richness, and so we conclude that the relations we find between radio luminosity and cluster richness are not driven by host-galaxy properties. Instead, this link is therefore likely to originate from a link between hot-gas fuelling and jet power.
\item We find a strong trend in AGN location within the group/cluster with radio luminosity: the lowest luminosity sources are likely to be at a large distance from the group/cluster centre, while the most radio-luminous AGN are typically close to the centre. The fraction of associated AGN hosted by a central galaxy also increases with radio luminosity. While host-galaxy properties do not appear to drive the previous reported trends between richness and luminosity, stellar mass is likely to be the dominant cause of the different location preferences of low and high luminosity AGN.
\item We find significant differences in the environmental properties of FRI and FRII radio galaxies, consistent with previous work: FRI radio galaxies show a systematic increase in cluster association fraction with radio luminosity, and a systematically higher association fraction than FRIIs at high luminosity. Similarly we find evidence that at high radio luminosities, FRI radio galaxies inhabit systematically richer environments than FRIIs. 
\end{itemize}
 
 The results presented here will provide useful input for numerical models of radio-galaxy evolution and modelling of the radio-loud AGN population and its evolution, which so far have not incorporated a link between jet power/radio luminosity and environmental richness \citep[e.g.][]{turner15,hardcastle18a, hardcastle18b}. These relationships also provide important observational constraints with which cosmological models of galaxy and cluster evolution should ensure consistency. With deeper redshift information and upcoming larger sky areas for LoTSS, and the use of deeper optical and infrared data, in future we will be able to build on this low-redshift baseline to explore the relationship between AGN jet populations and large-scale environments at earlier epochs where feedback mechanisms are less well understood. 
 
%\begin{figure*}
%   \centering
%   \includegraphics[width=8cm]{det_lambda.pdf}
%   \includegraphics[width=8cm]{det_rl.pdf}
   %%%\includegraphics{empty.eps}
   %%%\includegraphics{empty.eps}
%   \caption{AGN detection fraction as a function of richness, using the RedMaPPer $\lambda$ richness estimator (left) and using the W12 $RL\_$ estimator (right).}
%              \label{det_rich}%
%    \end{figure*}
%

%__________________________________________________ One column table
   
%

\section*{Acknowledgements}

% author acknowledgements
JHC and BM acknowledge support from the Science and Technology Facilities Council (STFC) under grants ST/R00109X/1 and ST/R000794/1. MJH and WLW acknowledge support from the UK Science and Technology Facilities Council (STFC) [ST/M001008/1]. PNB and JS are grateful for support from the UK STFC via grant ST/M001229/1. KJD acknowledges support from the ERC Advanced Investigator programme NewClusters 321271. GG acknowledges a CSIRO OCE Postdoctoral Fellowship. MB and IP acknowledge support from INAF under PRIN SKA/CTA ‘FORECaST’. LKM acknowledges support from Oxford Hintze Centre for Astrophysical Surveys which is funded through generous support from the Hintze Family Charitable Foundation. This publication arises from research partly funded by the John Fell Oxford University Press (OUP) Research Fund. SPO acknowledges financial support from the Deutsche Forschungsgemeinschaft (DFG) under grant BR2026/23. 

LOFAR, the Low Frequency Array designed and constructed by ASTRON, has facilities in several countries, which are owned by various parties (each with their own funding sources), and are collectively operated by the International LOFAR Telescope (ILT) foundation under a joint scientific policy. The ILT resources have benefited from the following recent major funding sources: CNRS-INSU, Observatoire de Paris and Université d’Orléans, France; BMBF, MIWF-NRW, MPG, Germany; Science Foundation Ireland (SFI), Department of Business, Enterprise and Innovation (DBEI), Ireland; NWO, The Netherlands; the Science and Technology Facilities Council, UK; Ministry of Science and Higher Education, Poland. Part of this work was carried out on the Dutch national e-infrastructure with the support of the SURF Cooperative through grant e-infra 160022 \& 160152. The LOFAR software and dedicated reduction packages on \url{https://github.com/apmechev/GRID_LRT} were deployed on the e-infrastructure by the LOFAR e-infragroup, consisting of J. B. R. Oonk (ASTRON \& Leiden Observatory), A. P. Mechev (Leiden Observatory) and T. Shimwell (ASTRON) with support from N. Danezi (SURFsara) and C. Schrijvers (SURFsara). This research has made use of the University of Hertfordshire high-performance computing facility (\url{http://uhhpc.herts.ac.uk/}) and the LOFAR-UK computing facility located at the University of Hertfordshire and supported by STFC [ST/P000096/1]. This research made use of Astropy, a community-developed core Python package for astronomy (Astropy Collaboration et al. 2013) hosted at http://www.astropy.org/, of Matplotlib (Hunter 2007), and of topcat (Taylor 2005). The Pan-STARRS1 Surveys (PS1) have been made possible through contributions by the Institute for Astronomy, the University of Hawaii, the Pan-STARRS Project Office, the Max-Planck Society and its participating institutes, the Max Planck Institute for Astronomy, Heidelberg and the Max Planck Institute for Extraterrestrial Physics, Garching, The Johns Hopkins University, Durham University, the University of Edinburgh, the Queen’s University Belfast, the Harvard-Smithsonian Center for Astrophysics, the Las Cumbres Observatory Global Telescope Network Incorporated, the National Central University of Taiwan, the Space Telescope Science Institute, and the National Aeronautics and Space Administration under Grant No. NNX08AR22G issued through the Planetary Science Division of the NASA Science Mission Directorate, the National Science Foundation Grant No. AST-1238877, the University of Maryland, Eotvos Lorand University (ELTE), and the Los Alamos National Laboratory. Funding for SDSSIII has been provided by the Alfred P. Sloan Foundation, the Participating Institutions, the National Science Foundation, and the U.S. Department of Energy Office of Science. The SDSS-III web site is \url{http://www.sdss3.org/}. SDSSIII is managed by the Astrophysical Research Consortium for the Participating Institutions of the SDSS-III Collaboration including the University of Arizona, the Brazilian Participation Group, Brookhaven National Laboratory, Carnegie Mellon University, University of Florida, the French Participation Group, the German Participation Group, Harvard University, the Instituto de Astrofisica de Canarias, the Michigan State/Notre Dame/JINA Participation Group, Johns Hopkins University, Lawrence Berkeley National Laboratory, Max Planck Institute for Astrophysics, Max Planck Institute for Extraterrestrial Physics, New Mexico State University, New York University, Ohio State University, Pennsylvania State University, University of Portsmouth, Princeton University, the Spanish Participation Group, University of Tokyo, University of Utah, Vanderbilt University, University of Virginia, University of Washington, and Yale University. This publication makes use of data products from the Wide- field Infrared Survey Explorer, which is a joint project of the University of California, Los Angeles, and the Jet Propulsion Laboratory/California Institute of Technology, and NEOWISE, which is a project of the Jet Propulsion Laboratory/California Institute of Technology. WISE and NEOWISE are funded by the National Aeronautics and Space Administration.

\bibliographystyle{aa}
\bibliography{qjet}

\end{document}